\documentclass[showpacs,prd,twocolumn,amsmath,amssymb]{revtex4-1}
\usepackage{graphicx}% Include figure files
\usepackage{amsmath,amsfonts}
\usepackage{dcolumn}% Align table columns on decimal point
\usepackage{bm}% bold math
\usepackage{slashed}
\usepackage{nicefrac}
\usepackage{epstopdf}
\usepackage{color}
\usepackage{tikz-feynman,contour}

\def\Tr{\mathrm{Tr}}
\def\Sp{\mathrm{Sp}}
\def\cD{\mathcal{D}}

\def\cA{\mathcal{A}}

\def\R{\mathbb{R}}
\def\Ei{\mathrm{Ei}}
\def\dk#1#2{\frac{ d^{#2}{#1} }{ (2\pi)^{#2} }} % invariant measure in FT
\begin{document}
\title{Wavelet regularization of Euclidean QED}
\author{Mikhail V. Altaisky}
\affiliation{Space Research Institute RAS, Profsoyuznaya 84/32, Moscow, 117997, Russia}
\email{altaisky@rssi.ru}
\author{Robin Raj}
\affiliation{Mahatma Gandhi University,  Priyadarsini Hills, Kottayam, Kerala, 686560, India}
\email{robuka97@gmail.com}
\date{Revised Nov 7, 2020} % It is always \today, today,
\begin{abstract}
The regularization of  quantum electrodynamics in the space of functions 
$\psi_a(x)$, which depend on both the position $x$ and the scale $a$, is presented. 
The scale-dependent functions are defined in terms of the continuous wavelet transform 
in $\R^4$ Euclidean space, with the derivatives of Gaussian served as mother wavelets.
The vacuum polarization and the dependence of the effective coupling constant on 
the scale parameters are calculated in one-loop approximation in the limit $p^2 \gg 4m^2$.
\end{abstract}

\maketitle

\section{Introduction \label{intro:sec}}
This paper was initially conceived as an erratum to the paper \cite{AK2013}, where 
we have found  technical errors in the evaluation of one-loop diagrams [Eqs.(34,36)] in wavelet-based quantum electrodynamics (QED). However, it was found later, that a
simple model of wavelet-based QED, briefly described in the aforementioned paper, can 
shed some new light on the scale dependence of the coupling constant on the observation 
scale in an Abelian gauge theory -- starting from completely finite quantum field theory 
model with no need of renormalization.  

In the previous papers \cite{AK2013,Altaisky2020PRD}, the possibility to construct a 
finite theory of scale-dependent fields $\psi_a(x)$ was developed, where the field $\psi_a(x)$
describes the fluctuations of typical size $a$. In this paper, we make a simplifying assumption that 
all measurable quantities can be determined in terms of effective fields $\psi_A(x) \sim \sum_{A\le a \le \infty}\psi_a(x)$ (with the meaning of the sum clarified later in the text), which are the sums of all 
fluctuations larger than the observation scale $A$. This approach allows us to start with a standard 
QED Lagrangian at large scales, with the ''bare'' coupling constant understood as a physical electron 
charge $\frac{e_0^2}{4\pi}\approx \frac{1}{137}$. In this sense, our approach of integrating from 
large scales to small scales is opposite to that used in standard RG calculations 
  \cite{Collins1997},
where the bare charge is formally located at infinitely small scales. The physical results at any finite 
observation scale, of course, should not depend on the direction in which we sum the fluctuations of different 
scales.

%%%%%%%%%%%%%%%%%%% Rev Oct 31 %%%%%%%%%%%%%%%%%%
Wavelets have been entering different branches of physics since the late 1980-s as an efficient tool 
for data processing. The main idea of wavelet transform is to unfold a function, an image, or other types 
of data gradually, scale-by-scale, from the coarsest scale to the finest details. In some sense, the idea 
of wavelet transform is reverse to the renormalization procedure, Kadanoff's blocking, etc. Renormalization 
gradually integrates out the details of finer scales in order to obtain effective interaction at a coarser scale. 
The wavelet representation gradually adds the details of finer scales to reconstruct the detailed picture starting 
from the coarsest snapshot. Both procedures are essentially related to the idea of self-similarity, which can be 
easily implemented in Euclidean space. That is why, most known applications of wavelets to quantum field theory problems 
deal either with the lattice regularization or with the Euclidean versions of QFT models \cite{Battle1999,Best2000}.
More recent applications of wavelets to lattice theories are related to the so-called multiscale entanglement 
renormalization (MERA)\cite{SMB2018}; some recent results can be found in \cite{Brennen2015,Fries2019}. 

Naively, one could expect that the results obtained with wavelets in Euclidean QFT models can be 
analytically continued to the Minkowski space. This is not so straightforward, since the introduction of a 
new scale variable implies the ordering of field operators in both the time and the scale arguments. 
The construction of consistent theory directly in Minkowski space still remains an open problem. Most likely, 
the solution of this problem can be found using the light-front coordinates, as suggested in 
\cite{PG2012,AK2013iv,AK2016IJTP,Polyzou2020}. The research in this direction is going on but is not 
a subject of this paper. Since this is not done yet, we study quantum field theory problems in Euclidean settings.
%%%%%%%%%%%%%%%%%%%%%%%%%%%%%%%%%%%%%%%%%%%%%%%%%

The remainder of this paper is organized as follows: In {\em Sec.~\ref{qed:sec}}, we summarize the 
scale-dependent approach to QED, described in the previous paper \cite{AK2013}, and present the results 
of one-loop calculations performed in Euclidean $\R^4$ space, with two different wavelets, {\sl viz.} the 
first and the second derivatives of the Gaussian. {\em Sec.~\ref{ward:sec}} accounts for the role of gauge invariance and  corresponding Ward-Takahashi identities, which stem from 
this invariance. We have shown by direct calculation that in the theory with local gauge invariance, 
$\psi(x)\to e^{-\imath e \Lambda(x)}\psi(x)$, defined for local fields, the Ward identity 
$\partial_\mu \Pi_{\mu\nu}=0$ is violated for any finite scale $A>0$. 
In {\em Conclusions}, we summarize the reasons for violation 
of a locally-defined gauge invariance by finite-scale wavelet calculations and propose to substitute it 
by the scale-dependent gauge invariance, that has been already considered by different authors 
\cite{Gu2006,Altaisky2020PRD}. 

\section{Wavelet-based regularization in quantum electrodynamics \label{qed:sec}}
Quantum electrodynamics  was the first quantum field theory model to face the 
problem of deriving finite observable quantities -- physical charge and physical mass of the electron -- from formally divergent Feynman integrals. Formal solution of this problem has been found in terms of the renormalization group (RG) formalism \cite{SP1953,Bsh1956}, which is 
physically related to the assumption of self-similarity of underlying physical processes \cite{WK1974}.   The renormalization procedure consists of two steps. The first step is 
the {\em regularization} -- formal subtraction of the divergent parts of Feynman integrals. The second step is the {\em multiplicative renormalization} of the fields and the model parameters so that the theory of new (renormalized) fields becomes 
finite. Different technical means of regularization have been proposed, see, e.g. \cite{Polchinski1984,BTW2002}. Most of them 
are essentially based on subtracting infinities from the Green functions defined in 
a space of square-integrable functions of either Minkowski or Euclidean coordinate.

However, there is an alternative point of view on the divergences in quantum field theory \cite{Altaisky2010PRD}. An attempt to measure any physical field {\em sharp} at a point $x$, with an 
infinite resolution $a\to0$, inevitably demands an infinite energy injection with a momentum of order $\frac{\hbar}{a}$, which would certainly destroy the system to be measured. This makes the pointwise definition of fields physically meaningless. As it concerns phenomenology, the initial and the final states of particles in 
high-energy physics experiments are usually determined in momentum space, i.e., in the basis of plane waves. 
For this reason, the results of measurements are considered as functions of different form factors, dependent on squared  momentum transfer $Q^2$ \cite{DEUR2016}. $Q^2$ partially plays the role of an observation scale, but this 
approach cannot completely reveal the spatial structure of interactions. This is because the Fourier transform, 
being based on the group of translations, is non-local, and the study of the $Q^2$-dependence does not allow 
for revealing of local details.  

There is a counterpart of such incompleteness in classical physics. Suppose we have a system with two high-frequency 
harmonics $\omega_1$ and $\omega_2$, the difference between which is a low-frequency harmonic: 
$\Delta\omega=|\omega_1-\omega_2| \ll \frac{\omega_1 + \omega_2}{2}$. Measuring the spectrum of such system 
we can observe a low-frequency harmonic $\Delta\omega$, but we cannot be sure, whether it originates from the existing 
large-scale structures, or it is just an artefact of beating between the two high-frequency harmonics -- unless we 
extend the frequency measurements ($\omega$) to frequency-scale measurements ($\omega,a$), and find out whether our 
observation comes from large $a$  or from small $a$ values. This method is often used in geophysics \cite{GGM1984}. 

By analogy, we think a phenomenologically consistent description of physical fields 
should incorporate both the position ($x$) and the scale ($a$). The more parameters we have the more detailed 
information we can get.

%%%%%%%%%%%%%%%%%%%% Rev Oct 30 %%%

Technical way to the construction of quantum field theory models for the fields 
$\psi_a(x)$ that depend on both the coordinate and the scale (resolution) from 
very beginning is provided by continuous wavelet transform \cite{Altaisky2010PRD,AK2013}. The scale-dependent Green functions $\langle \psi_{a_1}(x_1) \ldots\psi_{a_n}(x_n) \rangle$ are finite by construction. 

The simplest way to construct a field theory for the scale-dependent fields 
$\psi_a(x)$ is to express the  fields $\psi(x) \in \mathrm{L}^2(\R^d)$ in terms 
of their wavelet transform
\begin{equation}
\psi(x) = \frac{1}{C_\chi} \int_{\R_+ \otimes \R^d}
\frac{1}{a^d} \chi \left(\frac{x-b}{a} \right) \psi_a(b)\frac{dad^db}{a}
\label{iwt}
\end{equation}
in the original quantum field theory model built for the fields $\psi(x)$. 
%%%%%%%%%%%%%%% Rev Oct 29 %%%%%%%%%%%%%%%%%%%%5
Here $\chi(x)$ is some well-localized function (see, e.g, \cite{Daub10} for more details of the wavelet transform), usually referred to as a {\em mother wavelet}, or a {\em basic wavelet}. 
$C_\chi$ is the normalization constant defined below. 
%%%%%%%%%%%%%%%%%%%%%%%%%%%%%%%%%%%%%%%%%%%%%%%%%%%%%%% 
The coefficients 
\begin{equation}
\psi_a(b):= \int_{\R^d} \frac{1}{a^d} \bar{\chi}\left(\frac{x-b}{a} \right) \psi(x)d^dx  \label{cwt}
\end{equation}
are known as {\em wavelet coefficients} of $\psi$ with respect to the {\em mother wavelet} $\chi$. In fact, the transform \eqref{iwt} is a particular case of the 
{\em partition of unity} with respect to square-integrable representation 
$U(g), g\in G$ of a Lie group $G$:
$$
\hat{1} = \frac{1}{C_\chi} \int_G U(g) |\chi\rangle d\mu_L(g) \langle \psi| U^\dagger (g),
$$
for the case of $G$ being the affine group $G:x'=ax+b, a\in \R_+, b,x \in \R^d$ \cite{DM1976}. Here we have simplified the matter assuming the basic wavelet $\chi$ 
to be {\em isotropic}, and exclude $SO(d)$ rotations from the left-invariant measure $d\mu_L(g)$ on the Lie group $G$.

%%%%%%%%%%% Rev Oct 29 %%%%%%%%%%%%%%%%
For an isotropic wavelet $\chi$ the sufficient condition to ensure that the wavelet transform \eqref{cwt}  is 
invertible and it's inverse \eqref{iwt} identically recovers the function $\psi(x)$,
%%%%%%%%%%%%%%%%%%%%%%%%%%%%%%%%%%%%%%%
 is a finite normalization of the basic wavelet $\chi$ with respect to the 
group of scale transformations, defined as:
\begin{equation}
C_\chi = \int_0^\infty |\tilde \chi(ak)|^2\frac{da}{a}
 < \infty.
\label{adcfi}
\end{equation}
Tilde means the Fourier transform: $\tilde{\chi}(k) = \int e^{\imath k x} \chi(x) d x$.
More details on continuous wavelet transform can be found in many monographs, e.g. in 
\cite{Daub10,Chui1992}. 

In common quantum field theory models, say in $\phi^4$ model, the field function $\phi(x)$ 
is a scalar product of the state vector of the field $|\phi\rangle$, and a state vector which corresponds to localization at the point $x$: $\phi(x):= \langle x | \phi \rangle$. Similarly, in wavelet-based theory 
$$
\psi_a(x) = \langle x,a; \chi | \psi \rangle,
$$
where the l.h.s of the scalar product corresponds to the settings of measurement, 
which can be potentially performed on the field $\psi$ by a device described by 
the {\em aperture function} $\chi$ -- this is an interpretation borrowed from optics 
\cite{PhysRevLett.64.745}. The reason for the introduction of the parameters of observation $(\chi, a)$ into the definition of fields is a potential benefit of 
getting a field theory finite by construction.

%%%%%%%%%%%%%%%% Rev Oct 31 
Why should we use something else than the standard basis of plane waves? 
The basis of plane waves is the simplest basis used for analytical calculations in 
QED, and it is phenomenologically adequate to the registration of particles far from 
reaction domain. However, it is not the ultimate one. 
Depending on the symmetry of the problem, some other bases may be used to effectuate the calculations.
For the symmetry reason, considering the QED of an atom near a curved metallic surface the calculations 
can be performed in a basis of spherical functions \cite{Hetet2010}.  

In high-energy physics experiments the detectors are far from the reaction centre, and 
there is no need to look for  localized solutions to effectuate the calculations. On the 
other hand, the calculations in the basis of plane waves suffer from formal divergences, 
and for this reason, since the restrictions on the basic wavelet $\chi$ in \eqref{iwt} are 
very loose, we can attempt to use a localized basis to find a better solution than the standard one.
 
By analogy with optics, we can expect that the best basic function would be the aperture function 
of measuring device \cite{PhysRevLett.64.745}, but such functions are not feasible for analytical 
calculations. For this reason, we have either to use some simple function, which enables for analytical 
calculations, and in some sense resembles the aperture, or to do the calculations numerically. 
    
Clearly, it still remains practically unfeasible to use a real aperture function 
of a physical device in analytical calculations. For this reason, we have to use some 
simple localized functions, satisfying the admissibility condition \eqref{adcfi}, as 
a mother wavelet in our calculations. Alternatively, the use of (discrete) wavelet transform in gauge theories 
have been first proposed in the context of QCD \cite{Federbush1995}, but have not 
succeeded for a number of reasons. First, the wavelet transform is a linear integral transform. Hence, it respects the 
linearity of the gradient transform of gauge fields in the Abelian gauge theory, but does not behave so for non-Abelian (i.e., nonlinear) gauge theories. Second, 
the linearity of wavelet transform imposes a question of whether we can 
respect the local gauge invariance of the matter fields: $\psi(x) \to e^{-\imath \alpha(x)} \psi(x)$. This question is partially discussed in \cite{Altaisky2020PRD}.
Third, the introduction of the scale argument into the definition of quantum fields 
imposes two types of causality conditions: the standard (signal) causality, which 
provides the time-ordering in Minkowski space, and the causality between the small and 
the large scales (the part -- the whole relations) \cite{CC2005,Sorkin2003,AltaiskyPEPAN2005}.   
Of course, this does not preclude either to use discrete wavelet transform with the summation 
over a discrete set of scales \cite{Battle1999,Polyzou2017} or to combine wavelet transform 
with light-front variables, which seems better from the standpoint of causality \citep{AK2016IJTP,Polyzou2020}.

The results obtained with different basic wavelets 
may be different from each other -- same as the pictures in optical microscopy obtained with different 
apertures. The invariants, such as the total current, should be the same. This is rather similar to 
standard calculations, where we have to integrate over the momentum range of the detector to estimate 
the probability of particle detection. In the case of wavelets, one should perform the integration in both 
the momentum range and the scale range, which depend on the chosen basic wavelet.

We skip these difficult questions now (but keep them for future research), and will 
concentrate on the Euclidean model, where the scale parameter, considered in Euclidean space, is merely the best attainable resolution. In this way, we assume that 
''physical'' fields are sums of all scale components up to the best resolution $\cA$:
\begin{equation}
\psi^{(\cA)}(x) = \frac{1}{C_\chi} \int_{a\ge \cA} \chi\left(\frac{x-b}{a} \right)
\psi_a(b) \frac{dad^b}{a}.
\end{equation} 
In this sense, wavelet-based regularization in quantum field theory is similar to 
the momentum cutoff $\Lambda$, but has an advantage of respecting translation 
invariance and the momentum conservation of each vertex of the Feynman diagrams.

We start with the (Euclidean) QED Lagrangian 
\begin{align}
L_E = \bar\psi(x)(\slashed{D}+ \imath m)\psi(x) + \frac{1}{4} F_{\mu\nu}F_{\mu\nu}+ \underbrace{\frac{1}{2\alpha} (\partial_\mu A_\mu)^2}_{\hbox{gauge fixing}},\label{gau1l}  \\ 
\nonumber 
\hbox{where\ }  F_{\mu\nu}=\partial_\mu A_\nu - \partial_\nu A_\mu, \quad 
D_\mu = \partial_\mu + \imath e A_\mu,
\end{align}
%%% Rev Oct 29
and $\alpha$ is the gauge-fixing parameter,
%%%%%%%%%%%%%%%%%%%%%%%%%%%% 
with the Euclidean gamma matrices obeying the anticommutation relation 
\begin{equation}
\gamma_\mu \gamma_\nu + \gamma_\nu \gamma_\mu = -2 \delta_{\mu\nu} \label{gacom}
\end{equation} 
in $d=4$ dimensions. Slashed vectors mean the convolution with the Dirac gamma matrices: $\slashed{D}\equiv \gamma_\mu D_\mu$. In this paper we use Euclidean notation, so all indices are the subscripts 
: $ab\equiv a_\mu b_\mu$.

The generating functional of the quantum field theory model 
\begin{align}\nonumber 
Z_E[J,\eta,\bar{\eta}] = \int \cD A \cD \bar{\psi} \cD\psi 
\exp \Bigg[
- \int L_E d^dx - \\ 
- \int \bigl( J_\mu(x) A_\mu(x) +\imath \bar{\psi}(x)\eta(x) +\imath \bar{\eta}(x)\psi(x) \bigr) d^dx
\Bigg]\label{ZE}
\end{align}
%%%%%%%%%%% Rev Oct 29 %%%%%%%%%%%%%%%%%%
(where $\eta(x)$ and $\bar{\eta}(x)$ are formal Grassman-valued source fields, and $J_\mu(x)$ is a formal vector 
source corresponding to electromagnetic field), 
%%%%%%%%%%%%%%%%%%%%
can be made into the generating functional for the scale-dependent fields ($A_{\mu,a}(x),\bar{\psi}_a(x),\psi_a(x)$)  by the expression of the 
original fields in terms of \eqref{iwt}. This gives:
\begin{align}\nonumber 
Z_W[J_a,\eta_a,\bar{\eta}_a] = \int \cD A_a \cD \bar{\psi}_a \cD\psi_a \\
\exp \Bigl(
- S_W[A_a,\bar{\psi}_a,\psi_a] 
- \int J_{\mu,a}(x) A_{a,\mu}(x) \frac{d^dx da}{C_\chi a}  \\
\nonumber 
- \imath \int \bar{\psi}_a(x)\eta_a(x)  \frac{d^dx da}{C_\chi a}
- \imath \int \bar{\eta}_a(x)\psi_a(x)  \frac{d^dx da}{C_\chi a}
\Bigr),
\end{align}
where the ''action functional'' $S_W[A_a,\bar{\psi}_a,\psi_a]$ is a nonlocal functional obtained 
by substitution of \eqref{iwt} into Euclidean action functional $S_E=\int L_E d^dx$, see \cite{AK2013} for details. 

This substitution takes the most simple form in Fourier representation, where the 
convolutions become products. 
%%%%%%%%%%%% Rev Oct 29 
In momentum space, the inverse wavelet transform \eqref{iwt} for any field $\psi$ becomes:
\begin{equation}
\psi(x) = \frac{1}{C_\chi} \int_0^\infty \frac{da}{a}\int \frac{d^dk}{(2\pi)^d}
e^{-\imath k x} \tilde{\chi}(ak) \tilde{\psi}_a(k), \label{iwtA}
\end{equation}
where 
\begin{equation}
\tilde{\psi}_a(k) = \overline{\tilde{\chi}(ak)}\tilde{\psi}(k) \label{cwtB}
\end{equation}
is wavelet image of the field $\psi$ written in Fourier representation.
%%%%%%%%%%%%%%%%%%%%%%%%%%%%%%%%%%%%%%%%%
 The relations 
(\ref{iwtA},\ref{cwtB}) provide a set of simple rules for building Feynman diagrams 
for scale-dependent fields \cite{Altaisky2010PRD}:
\begin{itemize}\itemsep=0pt
\item each field $\tilde\psi(k)$ will be substituted by the ''scale component'' \eqref{cwtB}:
$\tilde\psi(k)\to\tilde\psi_a(k) = \overline{\tilde \chi(ak)}\tilde\psi(k)$.
\item each integration in momentum variable is accompanied by corresponding 
scale integration:
\[
 \dk{k}{d} \to  \dk{k}{d} \frac{da}{a} \frac{1}{C_\chi}
 \]
\item each interaction vertex is substituted by its wavelet transform; 
for the $N$-th power local interaction vertex this gives multiplication 
by factor 
$\displaystyle{\prod_{i=1}^N \overline{\tilde \chi(a_ik_i)}}$.
\end{itemize}
This means we have changed the coordinates $\bm{x}$ [or $\bm{p}$] on the translation group to the coordinates ($\bm{x},a$) [or ($\bm{p},a$)] on the affine group and we go 
on with the integration over the left-invariant measure on the affine group.

Since the Eq.\eqref{iwtA} contains the integration in full range of scales $\int_0^\infty \frac{da}{a}$, providing an identity \eqref{iwt} by doing so, the integration over {\em all} scale arguments in infinite limits would certainly drive us back to the 
common divergent theory.

Here is a point to make some physical assumptions. If we admit, that our hypothetical 
equipment has the best resolution scale $\cA$, -- which corresponds to the minimal of all scales of the external lines of a Feynman diagram of a process we are going to measure, -- then the integration over the scale arguments of all {\em internal} lines will 
be restricted to the range $\int_\cA^\infty$. This is an assumption that the modes of  scales smaller than the best resolution are not excited \cite{Alt2002G24J}. It makes all Feynman diagrams integrated in this way UV-finite.

In Euclidean QED we have the following elements of Feynman diagrams:
 
propagator of the spin-half fermion: 
\begin{equation}\nonumber 
\feynmandiagram [horizontal=c to d] { 
    c -- [fermion,edge label=p] d,
  };= \tilde{\chi}(ap) \frac{\imath (\slashed{p}-m)}{p^2+m^2} \tilde{\chi}(-ap), 
\end{equation}

photon propagator (taken in Feynman's gauge): 
\begin{equation}\nonumber 
\feynmandiagram [horizontal=A to B] { 
    A -- [photon,momentum=p] B,
  };= \tilde{\chi}(ap)\frac{\delta_{\mu\nu}}{p^2}  \tilde{\chi}(-ap), 
\end{equation}

fermion-photon vertex:
\begin{equation}\nonumber 
\begin{tikzpicture}[baseline=(a)]
\begin{feynman}[horizontal = (a) to (b)]
\vertex (a);
\vertex [above left =1cm of a] (c) ;
\vertex [below left =1cm of a] (d) ;
\vertex [right = 1cm of a] (b) {\(\mu\)};
\diagram*{
(d) -- [fermion] (a) -- [fermion] (c),
(a) -- [photon] (b),
};
\end{feynman}
\end{tikzpicture}
= -\imath e \gamma_\mu \prod_{i=1}^3 \overline{\tilde{\chi}(a_ip_i)}.
\end{equation}
Since each  internal line in a Feynman diagram is connected to two vertexes, from the left and from the right, the integration in the left and the right scale 
arguments, according to the above imposed scale-limitation rule, results in a multiplier 
\begin{equation}
\int_\cA^\infty \frac{|\tilde{\chi}(a_Lk)|^2}{a_L C_\chi}da_L 
\int_\cA^\infty \frac{|\tilde{\chi}(-a_Rk)|^2}{a_R C_\chi}da_R = 
f^2(\cA k) , \label{cf}
\end{equation}
where 
\begin{equation}
f(x) = \frac{1}{C_\chi} \int_x^\infty \frac{\tilde{\chi}(a)}{a}da. \label{fch}
\end{equation}
is the {\em wavelet cutoff function}, which satisfies an evident condition $f(0)=1$.
If we are not interested in how the fields of different scales $\bar{\psi}_a(x)$, $\psi_{a'}(x')$ and $A_{\mu,a''}(x)$ interact with each other but are interested only in the total effect of all 
fluctuations of scales larger than $a$, we can merely insert the wavelet cutoff  factors in 
all internal lines of Feynman diagrams. 

\begin{figure}[ht]
\centering \includegraphics[width=8cm]{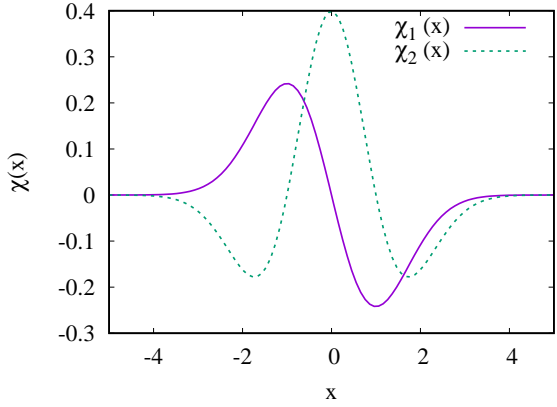}
\caption{First two wavelets of the Gaussian wavelet family \eqref{gn}}
\label{g12:pic}
\end{figure}

In our calculations, we use different derivatives of the Gaussian as mother wavelets. The admissibility condition \eqref{adcfi} is rather loose: practically any well-localized function with the Fourier image vanishing at zero momentum $\tilde{\chi}(0)=0$ obey this requirement. As for the Gaussian functions 
\begin{equation}
\chi_n(x) = (-1)^{n+1} \frac{d^n}{dx^n} \frac{e^-\frac{x^2}{2}}{\sqrt{2\pi}},\quad n>0,
\label{gn}
\end{equation}
%%%%%%%%%%%%%%%%%%%%%%%% Rev Oct 29
where $x$ is a dimensionless argument,
%%%%%%%%%%%%%%%%%%%%%%%%%%%%%%%%%% 
they are easy to integrate in Feynman diagrams. The graphs of first two wavelets of the \eqref{gn} family, 
$$
\chi_1(x) = -x e^{-\frac{x^2}{2}}, \quad \chi_2(x) = (1-x^2) e^{-\frac{x^2}{2}},
$$
are shown in Fig.~\ref{g12:pic}. 
Their Fourier images are 
\begin{equation}
\tilde{\chi}_n(k) = - (\imath k)^n e^{-\frac{k^2}{2}}.  
\end{equation}
Respectively, the normalization constants and the wavelet cutoff functions are:
\begin{align*}
C_{\chi_n} = \frac{\Gamma(n)}{2}, \quad f_{\chi_n}(x) = \frac{\Gamma(n,x^2)}{\Gamma(n)},
\end{align*}
where $\Gamma(\cdot)$ is the Euler gamma function, and $\Gamma(\cdot,\cdot)$ is the incomplete gamma function. 
For the first two wavelets  the wavelet cutoff functions are:  
\begin{equation}
f_{\chi_1}(x) = e^{-x^2}, \label{fch1}\quad  f_{\chi_2}(x) = (1+x^2) e^{-x^2}. 
\end{equation}

\begin{figure}[ht]
\centering \includegraphics[width=5cm]{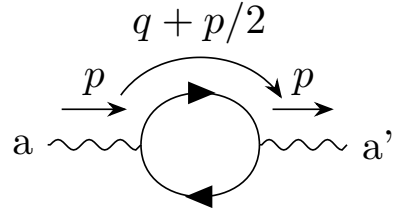}
\caption{Vacuum polarization diagram in scale-dependent QED}
\label{vpd:pic}
\end{figure}

\begin{figure}[ht]
\centering \includegraphics[width=5cm]{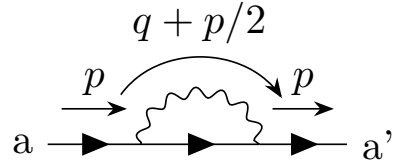}
\caption{Fermion self-energy diagram in scale-dependent QED}
\label{ese:pic}
\end{figure}

We will now proceed to the calculation of one-loop diagrams in wavelet-based 
Euclidean QED. These are the vacuum polarization 
diagram and the fermion self-energy diagram, shown in Figs.~\ref{vpd:pic},\ref{ese:pic}.

\paragraph{Vacuum polarization diagram.} First we calculate the vacuum polarization diagram shown in Fig.~\ref{vpd:pic}.
For the convenience of calculations, we symmetrize the loop momenta. The external lines 
of the diagram are labelled by scale arguments $a$ and $a'$. So, according to the  
assumptions made above, the integration in scale arguments in the fermion loop 
is limited from below by the minimal scale $\cA=\min(a,a')$. In contrast to the paper \cite{AK2013},
intended for calculation of the Green functions of scale-dependent fields $\langle \psi_{a_1}(x_1)\ldots \psi_{a_n}(x_n)\rangle$, here we do not specify any propagators 
on external lines so that the results can be taken as usual diagrams regularized 
to a scale $\cA$. That is why the wavelet factors are omitted in the definitions of 1PI diagrams. Doing so, we get the expression for the vacuum polarization diagram:
\begin{widetext}
\begin{align}
\Pi_{\mu\nu}^{(\cA)}(p)&=& -e^2 \int   
\frac{   \Sp (\gamma_\mu (\slashed{q}+ \frac{\slashed{p}}{2}  -m)\gamma_\nu (\slashed{q} -\frac{\slashed{p}}{2} - m))}{\left[(q+\frac{p}{2})^2+m^2\right]\left[(q-\frac{p}{2})^2+m^2\right]}  F_\cA(p,q) \dk{q}{4} \label{padef} \\
\nonumber &=& - 4e^2 \int   
\frac{2q_\mu q_\nu - \frac{1}{2} p_\mu p_\nu + 
\delta_{\mu\nu}(\frac{p^2}{4}-q^2-m^2)}{\left[(q+\frac{p}{2})^2+m^2\right]\left[(q-\frac{p}{2})^2+m^2\right]} F_\cA(p,q) \dk{q}{4}.
\end{align}
\end{widetext}
Here we use the function $Sp()$ to denote the trace of the Dirac gamma matrices.
The wavelet cutoff function $F_\cA(p,q)$ is the 
product of wavelet cutoff functions of the loop momenta: 
\begin{equation}
F_\cA(p,q) = f^2 \left(\cA \left(\frac{p}{2}-q\right) \right)f^2 \left(\cA \left(\frac{p}{2}+q\right) \right).
\end{equation}

Let us start the calculations with $\chi_1$ wavelet. In this case (Eq.\eqref{fch1}): 
$$
F_\cA(p,q)=e^{-\cA^2p^2 -4\cA^2q^2},$$
and we have the integral
\begin{align}\nonumber 
\Pi_{\mu\nu}^{(\cA,\chi_1)} &=& - \frac{e^2 p^2}{\pi^3} e^{-\cA^2p^2}\int_0^\infty 
dy y e^{-4\cA^2p^2y^2} \int_0^\pi d\theta \sin^{2}\theta \times \\
&\times& \frac{2y_\mu y_\nu - \frac{1}{2} \frac{p_\mu p_\nu}{p^2} + 
\delta_{\mu\nu}(\frac{1}{4}-y^2-\frac{m^2}{p^2})}
{\left[\frac{\frac{1}{4}+y^2+\frac{m^2}{p^2}}{y}+\cos\theta\right]
\left[\frac{\frac{1}{4}+y^2+\frac{m^2}{p^2}}{y}-\cos\theta\right]
}, \label{Pag1}
\end{align}
where we have introduced a dimensionless vector $y$ in the direction of loop momentum: 
$q = |p| y$, with $\theta$ being the angle between $p$ and $q$. The integral \eqref{Pag1} can be evaluated analytically 
in relativistic limit $p^2 \gg 4m^2$. This gives: 
\begin{align}
\Pi_{\mu\nu} &= I_T\left(\delta_{\mu\nu} - \frac{p_\mu p_\nu}{p^2}\right) 
+ I_L \frac{p_\mu p_\nu}{p^2}, \\   \nonumber 
I_T &= \frac{e^2p^2}{48\pi^2s^3}
\Bigl[
(4s^2-2s-1)e^{-2s} +(1+s-4s^2)e^{-s} \\   \nonumber 
&+ 4s^3 (\Ei_1(s)-2\Ei_1(2s))\Bigr], \\    \nonumber 
I_L &= \frac{e^2p^2}{16\pi^2 s^3} e^{-2s}\left((s-1)e^s+1\right),
\end{align}
where $s\equiv \cA^2p^2$ is dimensionless scale argument, $\Ei_1(z)=\int_1^\infty \frac{e^{-xz}}{x}dx$ is exponential integral of the first type. The details of the calculations 
are presented in the Appendix. 
%%% Rev. Nov 1 %%%%%%%%%%%%%%%%%%%%%%%
As we can see from  Eqs.(\ref{ID},\ref{IP}), the longitudinal part of $\Pi_{\mu\nu}$ does not vanish
in the limit of $s\to0$. In this sense, the wavelet observation 
scale $\cA$ plays the role of inverse regularising mass $\frac{1}{M}$ of the Pauli-Villars regularization \cite{Bsh1980}. 
%%%%%%%%%%%%%%%%%%%%%%%%%%%%%%%%%%%%%
In contrast to dimensional regularization, where $q_\mu q_\nu$ and $2q^2$ terms cancel 
each other in the sense of leading divergences, this does not happen in the theory with 
a finite scale $\cA$ and local gauge invariance. There may be different reasons for that. First, the finite terms, neglected by dimensional regularization turn into the scale-dependent contributions, which can't be neglected in our case. Second,  the scale $\cA$ is a scale in Euclidean space and we cannot match it exactly to what is measured in Minkowski space.  Third, changing the coordinates from $\mathbf{x}$ to $(\mathbf{x},a)$ we need to pay an extra attention on what is gauge invariance in scale-dependent 
settings \cite{Altaisky2020PRD} -- the consideration presented above ignored this completely by making standard assumption of local gauge invariance. 
\paragraph{Fermion self- diagram.}
The loop integral of the fermion self-energy diagram, shown in Fig.~\ref{ese:pic},
has the form:
\begin{equation}
\Sigma^{(\cA)}(p) = -\imath e^2 \int \dk{q}{4}  
\frac{F_\cA(p,q)
\gamma_\mu 
 \left[\frac{\slashed{p}}{2}-\slashed{q}-m \right] \gamma_\mu
 }
 {\left[
\left(\frac{p}{2}-q \right)^2+m^2\right]
 \left[\frac{p}{2}+q \right]^2
 }. \label{ese:def}
\end{equation}
As in the previous example, $A$ is the minimal scale of all external lines 
$\cA=\min(a,a')$.
We will calculate the diagram \eqref{ese:def} with the wavelet-cutoff functions $F_\cA(p,q)$ for both $\chi_1$ and $\chi_2$ wavelets (\ref{fch1}).

Using the identities for Euclidean gamma matrices, and assuming the relativistic limit 
$p^2\gg 4m^2$ for simplicity of calculations, we rewrite \eqref{ese:def} in the 
form
\begin{widetext}
\begin{equation}
\Sigma^{(\cA)}(p) = -\imath e^2 \int  \dk{y}{4} F_\cA(p,|p|y)
\frac{ (\slashed{p}+4m-2|p|\slashed{y})}
{
\left[y^2 + \frac{1}{4}-y\cos\theta -\frac{m^2}{p^2}\right]
\left[y^2 + \frac{1}{4}+y\cos\theta \right]
},\label{ese-y}
\end{equation} 
\end{widetext}
where the term proportional to $\slashed{y}$ in the numerator can be ignored, if
we make the denominator symmetric with respect to the inversions by omitting the mass term in fermion propagator. For the same reason of relativistic approximation $p^2\gg4m^2$, we can regard the mass term in the numerator as negligible in comparison to $\slashed{p}$. 

\begin{figure}[ht]
\centering \includegraphics[width=4cm]{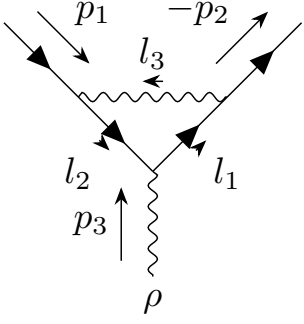}
\caption{One-loop contribution to the fermion-photon vertex in QED.  $p_1+p_2+p_3=0$.}
\label{epv:pic}
\end{figure}

Under the above-made assumptions, taking the wavelet cutoff function of the type 
$\chi_1$, (Eq.\ref{fch1}), we can easily see that 
$$
\Sigma^{(\cA)}_{\chi_1}(p)=-\frac{\imath e^2 \slashed{p} e^{-s}}{16\pi^4}\int
\frac{dy y e^{-4sy^2} \sin^2\theta d\theta}{
\left(y+\frac{1}{4y} \right)^2-\cos^2\theta
} = - \frac{\imath e^2 e^{-s}}{4\pi^3}J \slashed{p},
$$
where the integral $J$ is given by \eqref{J-int}. Thus we get: 
\begin{equation}
\Sigma^{(\cA)}_{\chi_1}(p)=-\frac{\imath e^2}{16\pi^2}
\left[2\Ei_1(2s) - \Ei_1(s) - \frac{e^{-2s}}{s} + \frac{e^{-s}}{s} \right]
\slashed{p}. \label{se1}
\end{equation}
\paragraph{Fermion-photon vertex.}
The one-loop contribution to the fermion-photon vertex is shown 
in Fig.~\ref{epv:pic}.
Since the bare fermion-photon vertex is $-\imath e \gamma_\rho$, 
we similarly normalize the vertex function: 
%%%%%%%%%%%%%%%%%%%%% Rev Oct 29 %%%%%%%%%
\begin{widetext}
\begin{align} 
-\imath e \Gamma_\rho^{(\cA)}(p_1,p_2,p_3) = -\imath e \gamma_\rho + (-\imath e)^3 \int \frac{d^4l}{(2\pi)^4} \frac{
\gamma_\mu \imath (\slashed{l}_2+m) \gamma_\rho \imath (\slashed{l}_1+m)\gamma_\mu
}{(l_2^2+m^2)(l_1^2+m^2)l_3^2} f^2(\cA l_1) f^2(\cA l_2) f^2(\cA l_3),  \label{gr}
\end{align}
\end{widetext}
where $f(x)$ is the wavelet-cutoff function given by \eqref{fch}. To get rid of the 
angle dependence in the wavelet cutoff factors we have symmetrized the loop momenta: 
$$l_1 = l + \frac{p_3-p_2}{3},\quad l_2 = l + \frac{p_1-p_3}{3},\quad l_3 = l + \frac{p_2-p_1}{3}. $$
To calculate the one-loop contribution to the vertex let us consider the decay of a photon 
with momentum $p_3=p$ into a fermion-antifermion pair. This corresponds to the loop momenta
\begin{equation}
 l_1=l+\frac{p}{2},\quad l_2=l-\frac{p}{2}, \quad l_3=l.
\end{equation}
Considering the relativistic case $p^2 \gg 4m^2$ we can omit the mass terms. This gives 
\begin{align*} A_\rho &= \gamma_\mu \left(\slashed{l}-\frac{\slashed{p}}{2}\right)\gamma_\rho 
\left(\slashed{l}+\frac{\slashed{p}}{2}\right)\gamma_\mu \\ 
&=2 \left(\slashed{l}+\frac{\slashed{p}}{2}\right) \gamma_\rho \left(\slashed{l}-\frac{\slashed{p}}{2}\right).
\end{align*}
The one-loop contribution to the vertex then takes the form
\begin{equation}
\Lambda_\rho\left(-\frac{p}{2},-\frac{p}{2},p\right) = - e^2 \int \frac{d^4l}{(2\pi)^4} 
\frac{A_\rho F_\cA(p,l)}{\left(l-\frac{p}{2}\right)^2\left(l+\frac{p}{2}\right)^2l^2}. \label{loop1}
\end{equation}
The vertex wavelet cutoff factor is the product of three wavelet cutoff functions:
\begin{equation}
F_\cA(p,l) = f^2\left(\cA\left(l-\frac{p}{2}\right)\right) 
f^2\left(\cA\left(l+\frac{p}{2}\right)\right) f^2(\cA l). 
\end{equation}
For the case of $\chi_1$ wavelet, see Eq.~\ref{fch1}, we have 
$$
F_\cA(p,l) = \exp\left( -\cA^2p^2 -6\cA^2l^2\right).
$$
The calculation of the integral \eqref{loop1} with this cutoff function, presented in Appendix, gives: 
\begin{align}\nonumber 
\Lambda_\rho\left(-\frac{p}{2},-\frac{p}{2},p\right) 
= \frac{e^2 \gamma_\rho}{3\pi^2}\Bigg[ 
e^\frac{s}{2}\Ei_1(3s) - \frac{e^\frac{s}{2}\Ei_1\left(\frac{3s}{2} \right)}{2} 
\\ - \frac{e^{-\frac{5s}{2}}}{8s} + \frac{e^{-s}}{12s} 
-\frac{ 5 e^{-s}\Ei_1\left(\frac{3s}{2} \right)}{16}  + \frac{e^{-s}}{36s^2}
- \frac{e^{-\frac{5s}{2}}}{36s^2} 
 \Bigg].
 \label{Lambda1}
\end{align}
In terms of the fine structure constant $\alpha(s) = \frac{e^2(s)}{4\pi}$ the one-loop contribution 
to the QED vertex \eqref{Lambda1} can be cast in the form 
\begin{align}
\alpha(s) &= \alpha\left[1+ \frac{4}{3\pi}\alpha R(s) \right]^2, \label{alpha1}\\ \nonumber 
R(s)      &= 
e^\frac{s}{2}\Ei_1(3s) - \frac{e^\frac{s}{2}\Ei_1\left(\frac{3s}{2} \right)}{2} 
- \frac{e^{-\frac{5s}{2}}}{8s} +\\ 
&+ \frac{e^{-s}}{12s}   
-\frac{ 5 e^{-s}\Ei_1\left(\frac{3s}{2} \right)}{16}  + \frac{e^{-s}}{36s^2}
- \frac{e^{-\frac{5s}{2}}}{36s^2} .
\end{align}
The graph of the running coupling constant $\alpha(s)$, calculated according to the formula \eqref{alpha1}, is 
shown in Fig.~\ref{alpha1:pic} below.
\begin{figure}[ht]
\centering \includegraphics[width=8cm]{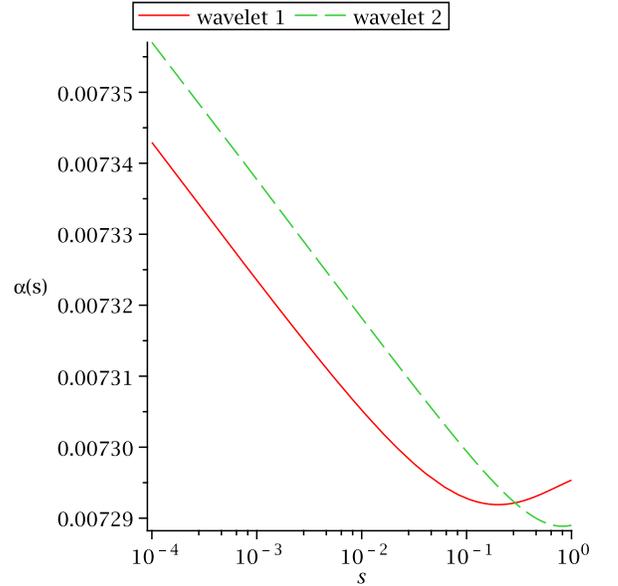}
\caption{Running coupling constant calculated for both wavelets $\chi_1$ and $\chi_2$ according to the formula \eqref{alpha1}. The value of the 
fine structure constant is $\alpha_\infty = 1/137.036$.}
\label{alpha1:pic}
\end{figure}
Decomposing Eq.\eqref{Lambda1} in a series for small scales ($s\to 0$)
$$
\Lambda_\rho \approx \gamma_\rho e^2 \frac{\frac{3}{2}-\frac{9}{16}\gamma + \frac{13}{16}\ln\frac{3}{2}-\frac{3}{16}\ln s-\ln3}{3\pi^2} + O(s),
$$
we get the logarithmic derivative 
\begin{equation}
\frac{\partial e(s)}{\partial \ln s} = -\frac{e^3}{16\pi^2}.
\end{equation}
The calculations performed with $\chi_2$ wavelet, presented in Appendix,
give similar results.
%which is in coincidence with the one loop correction to the QED vertex calculated by dimensional regularization 
%\begin{equation}
%\Gamma_\rho = \imath e \mu^\epsilon \gamma_\rho \frac{e^2}{16\pi^2} \frac{1}{\epsilon}.
%\end{equation}

\section{Ward identities \label{ward:sec}}
Formally, the Ward-Takahashi identities follow from a requirement that the Green function generating functional, 
designed on an action $S[\phi]$, should be invariant under the the same symmetry transformations $\phi(x) \to \phi(x) + \delta \phi(x)$ that 
leave the action invariant. For the case of spinor electrodynamics, the infinitesimal gauge transformations 
$\delta \phi$ take the form:
$$
\delta A_\mu = \partial_\mu \Lambda(x), \quad \delta\psi=-\imath e \Lambda(x)\psi, \quad 
\delta\bar{\psi} = \imath e \Lambda(x)\bar{\psi}.
$$
Since the  Lagrangian  is gauge-invariant by construction, to make the generation functional 
$Z[J,\bar{\eta},\eta]$ gauge invariant, we need to ensure that the variations of the source terms and the 
gauge-fixing terms compensate each other. This implies
\begin{align*}
\int \cD A \cD \bar{\psi} \cD \psi \left[e^{- \int d^4x(L_E + J_\mu A_\mu + \imath \bar{\eta}\psi +\imath \bar{\psi}\eta)}
e^{\delta_\Lambda}\right] = \\
=\int \cD A \cD \bar{\psi} \cD \psi e^{- \int d^4x(L_E + J_\mu A_\mu + \imath \bar{\eta}\psi +\imath \bar{\psi}\eta)},
\end{align*}
where 
\begin{equation}
\delta_\Lambda \equiv \int d^4x \left[
-\frac{1}{\alpha} \partial^2 (\partial_\mu A_\mu) + \partial_\mu J_\mu + e (\bar{\psi}\eta - \bar{\eta}\psi)
\right]\Lambda(x).
\end{equation}
Considering an infinitesimal transform we can approximate $e^{\delta_\Lambda}\approx 1+\delta_\Lambda$, and 
hence, in view of arbitrariness 
 of $\Lambda(x)$, the equality $\langle \delta_\Lambda \rangle=0$ can be written in a form of variational 
 equation:
 \begin{align} \label{wtig}
 \bigl[-\frac{1}{\alpha} \partial^2 (\partial_\mu A_\mu) + \partial_\mu J_\mu + e (\bar{\psi}\eta - \bar{\eta}\psi)
 \bigr]Z[J,\bar{\eta},\eta]=0, \\ \nonumber
 \hbox{where\ } \psi(x) = \imath\frac{\delta}{\delta \bar{\eta}(x)}, 
 \bar{\psi(x)} = \imath\frac{\delta}{\delta \eta(x)}, A_\mu = -\frac{\delta}{\delta J_\mu(x)} .
 \end{align}
 The Ward-Takahashi identities can be obtained by taking an appropriate number of functional derivatives of the equation \eqref{wtig}. This is usually done by changing from the generating functional $Z$ to the generating 
 functional for the connected Green's functions: 
 $$Z[J,\bar{\eta},\eta]= e^{-W [J,\bar{\eta},\eta]},$$
 and then applying the Legendre transform to get an effective action functional:
  $$\Gamma[A,\psi,\bar{\psi}] = W [J,\bar{\eta},\eta] - JA -\imath \bar{\eta}\psi - \imath\bar{\psi}\eta. $$
  The latter enables us to work with proper vertices and write the Ward-Takahashi identities generating equation in the form:
\begin{equation}
\frac{\partial^2}{\alpha} \partial_\mu A_\mu + \partial_\mu \frac{\delta \Gamma}{\delta A_\mu} 
+ \imath e \left(\psi \frac{\delta \Gamma}{\delta \psi} - \bar{\psi} \frac{\delta \Gamma}{\delta \bar{\psi}} 
\right)=0. \label{wige}
\end{equation}
%%%%%%%%%
The first derivative of \eqref{wige} with respect to $A_\mu$ gives the Ward identity \cite{Ward1950} that demands  
the transversality of the vacuum polarization diagram:
\begin{equation}
\partial_\mu \Pi_{\mu\nu} =0
\label{wti1}.
\end{equation}
The integrals in the vacuum polarization diagram will satisfy the requirement \eqref{wti1} only in case they are invariant under the shift 
of the loop momenta. This is not always true when a regularization procedure is applied. In QED, the condition 
\eqref{wti1} is observed by dimensional regularization, but not by the momentum cutoff. That is why dimensional 
regularization has become the most common regularization method in QFT models \cite{HooftVeltman1972}. 

To fulfil the Ward-Takahashi identities, a regulator is usually assumed to satisfy the requirement of the type \cite{BN1999,CL2011}:
\begin{equation}
\int \frac{d^4l}{(2\pi)^4} \frac{l_{\mu}l_{\nu}}{(l^2+\Delta)^2} = 
\frac{\delta_{\mu\nu}}{2}  \int \frac{d^4l}{(2\pi)^4} \frac{1}{l^2+\Delta}
\label{tr}
\end{equation} 
(the integration over the Feynman $x$-parameter used to get rid of angle integrations is not shown here).
This is definitely true for dimensional regularization, but is not 
true for momentum-cutoff and is not true for the wavelet regularization we consider in this paper. In our 
case of finite theory we cannot use a relation like \eqref{tr} as a ''rule'', but have to evaluate everything 
explicitly.

Regardless of the undoubted merits of dimensional regularization, it deals only with the main singular parts of 
Feynman diagrams, and cannot tackle the amplitudes at finite scales. In this respect, the finite 
cutoff regularization and the wavelet regularization have the potential advantage of describing of 
what happens at a finite observation scale \cite{Gu2002,Altaisky2010PRD}. The  goal of the wavelet-cutoff technique, provided by continuous wavelet transform, is to get a capability of calculations 
at finite observation scale.
%%%%% Rev Oct 29
 The respect to the gauge invariance can be also imposed in a cutoff-momentum regularization scheme by assuming the gauge 
transformations to act {\sl below} the momentum cutoff $\Lambda$ \cite{Gu2006}:
%%%%%%%%%%%%%%%
\begin{equation}
a_\mu(k) \to a_\mu(k) - \imath k_\mu \lambda(k), \quad \hbox{with\ } \lambda(|k|>\Lambda)=0.
\label{Gu}
\end{equation}
In the case of continuous wavelet transform regularization, there is an alternative -- to consider 
gauge transformations which directly depend on the scale argument $a$ :
\begin{equation}
\psi_a(x) \to e^{-\alpha_a(x)}\psi_a(x) \label{A101}
\end{equation}
Doing so, we get a theory that is gauge invariant separately at each given scale \cite{Altaisky2020PRD}.

We do not consider  scale-dependent modifications of gauge invariance in this paper, leaving 
this subject for future studies. Instead, the above considered wavelet cutoff factors of Gaussian type, are 
rather similar to already proposed exponential modifications of the momentum cutoff \cite{Oleszczuk1994}, based on 
the Schwinger proper time method \cite{Schwinger1951}.
Using wavelet regularization, in the case of small scales $s \ll 1$, when in the final limit of $s\to0$ the integration 
over {\em all} scales $\int_0^\infty \frac{da}{a}$ would definitely restore the symmetries of the original 
theory, we can use the approximation formulae that follow from Ward identities of the full (non-regularized) 
theory.
 
Technically, the Ward identities follow from the observation that a proper vertex of 
the fermion-photon interaction can be associated with the fermion self-energy 
diagram by inserting a photon line in the internal fermion line of the latter. 
Ward noticed, that for the bare inverse electron propagator
$$
S_{(e)}^{-1}(p) = \imath(\slashed{p}+m),
$$
the derivative with respect to the momentum $p_\mu$ gives the fermion-photon 
interaction vertex:
$$
\frac{\partial S_{(e)}^{-1}(p)}{\partial p_\mu} = \imath \gamma_\mu.
$$
He proved the same for the inverse full propagator: 
\begin{equation}
\frac{\partial G_{(e)}^{-1}(p)}{\partial p_\mu} = \imath \Gamma_\mu,
\end{equation}
where $-\imath e \gamma_\mu$ and  $-\imath e \Gamma_\mu$ are the bare and the 
full vertex of the fermion-photon interaction. (Here we use the Euclidean notation, in contrast to the original paper of Ward \cite{Ward1950}, written in Minkowski space.)

More generally, the Ward-Takahashi \cite{Takahashi1957} identity in spinor electrodynamics, 
written in integral form, 
relates the vertex function to the difference of fermion propagators:
\begin{equation}
q_\mu \Gamma_\mu(p,-p-q,q) = G^{-1}(p+q) - G^{-1}(p).
\label{wt1}
\end{equation}
Here $G(p)$ is the full fermion propagator. 
The identity 
\eqref{wt1} is a helpful constraint which ensures the gauge invariance 
of the renormalized QED in any order of perturbation theory 
\cite{Ward1950,Takahashi1957}. The constraint \eqref{wt1} makes 
the perturbation expansion gauge-invariant at the presence of the 
gauge fixing terms in the QED generating functional.

The most direct application of Ward's finding is the calculation of 
the full fermion-photon vertex in the limit of zero photon momentum. In this case:
\begin{equation}
\Gamma_\mu(p,-p,0) = \gamma_\mu + \Lambda_\mu(p,-p,0).
\end{equation}
As it follows from the Dyson equation, the inverse full propagator is equal to 
\begin{equation}
G^{-1}(p) = S^{-1}(p) - \Sigma(p), \label{dec}
\end{equation}
where $\Sigma(p)$ is the electron self-energy. Taking the derivatives of both sides 
of \eqref{dec} by $\frac{\partial}{\partial p_\mu}$ we get 
\begin{equation}
\Lambda_\mu(p,-p,0) = \imath \frac{\partial \Sigma(p)}{\partial p_\mu}. \label{WD}
\end{equation}

\begin{figure}[ht]
\centering \includegraphics[width=8cm]{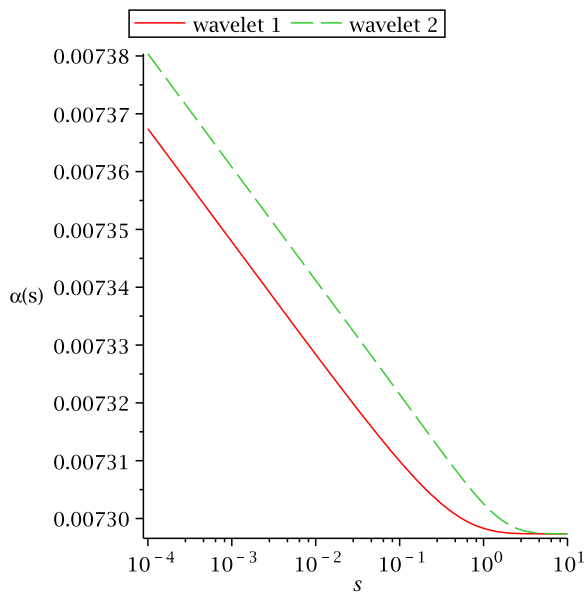}
\caption{Scale dependence of the fine structure constant 
$\alpha(s)=\frac{e_{eff}^2(s)}{4\pi}$. Two curves correspond to the one-loop approximation \eqref{rg2}, performed with 
$\chi_1$ and $\chi_2$ wavelet cutoff, respectively.}.
\label{sl:pic}
\end{figure}

The formula \eqref{WD} can now be applied to our wavelet-regularized calculations 
of one-loop diagrams. Since we are interested only in the contributions to the vertex, 
 proportional to $-\imath e \gamma_\mu$, it is sufficient to differentiate only the last multiplier in \eqref{se1}: $\frac{\partial\slashed{p}}{\partial p_\mu}=\gamma_\mu$.
This gives the one-loop equation for the fermion-photon vertex regularized at scale $\cA$:
\begin{align}\nonumber 
-\imath e\Gamma^{(\cA)}_{\chi_1,\mu}(p)&=&-\imath e \gamma_\mu \left[1+ \frac{e^2}{16\pi^2}
R_1^{\chi_1}(s) \right]+ \ldots , \\
R_1^{\chi_1}(s) &\equiv& 2\Ei_1(2s) - \Ei_1(s) - \frac{e^{-2s}}{s} + \frac{e^{-s}}{s}. 
\label{rg1}
\end{align}  
Since whole dependence on the scale in our model is contained in function $R_1(s)$, 
we can now calculate the dependence on the scale of the effective charge. The wavelet regularization scheme includes the 
integration of over all scale components from observation scale $A$ to infinity.
The equation \eqref{rg1} thus gives the value of the effective charge $e_{eff}(s)$, 
i.e. the effective charge measured at scale $\cA$, in terms of physical electron 
charge measured at infinity $e_0=e(\infty)$. It is convenient to rewrite it in terms of the fine structure constant 
$$
\alpha(s) = \frac{e^2(s)}{4\pi},
$$
the physical value of which is $\alpha\approx 1/137.036$ \cite{Dyck1987,KN2006}.
Then, the scale dependence of the effective charge, given in one-loop approximation by
the equation \eqref{rg1}, is:
\begin{equation}
\alpha(s) = \alpha\left(1+ \frac{\alpha}{4\pi}R_1(s) \right)^2.
\label{a1l}
\end{equation} 
 Since we use the coordinate scale $a$ as the scale argument 
the sign will be opposite to that in dimensional regularization $s\frac{\partial}{\partial s} \to - \mu^2 \frac{\partial}{\partial \mu^2}$.
The scaling equation for the effective charge -- we do not call it RG-equation, since 
there is no field renormalization in our model -- takes the form:
\begin{align}
s \frac{\partial e_{eff}}{\partial s} = \frac{e_{eff}^3}{16\pi^2} s \frac{\partial R_1(s)}{\partial s},\label{rge} \\ \nonumber 
 s \frac{\partial R_1(s)}{\partial s}=
\frac{e^{-s}}{s}\left(e^{-s}-1\right). 
\end{align}
The scaling equation \eqref{rge} can be integrated in a usual RG-like the form: 
\begin{equation}
\frac{d e_{eff}}{e_{eff}^3} = \frac{ds}{16\pi^2} \frac{e^{-s}}{s^2}
\left(e^{-s}-1\right). \label{rge1}
\end{equation}
The solution of the equation \eqref{rge1} is given by 
$$
e_{eff}^2(s) = \frac{e_0^2}{1 - \frac{e_0^2}{8\pi^2} R_1(s)},
$$
which can be cast in terms of the fine structure constant:
\begin{equation}
\alpha(s) = \frac{\alpha}{1 - \frac{\alpha}{2\pi} R_1(s)}.
\label{rg2}
\end{equation}
Similar results can be obtained using other wavelets. In this way using $\chi_2$ 
wavelet cutoff, see Eq.\eqref{sigma-chi2} in Appendix, we get:
\begin{align}\nonumber 
R_1^{\chi_2}(p) &= 2\Ei_1(2s) - \Ei_1(s)  -\frac{e^{-2s}(s+5)}{2s} +\\
&+ \frac{e^{-s}(s^3+18s^2+134s+640)}{256s}.
\end{align}
The scale dependences of $\alpha(s)$, calculated for both cases of \eqref{fch1}  wavelet 
cutoff functions, are shown in Fig.~\ref{sl:pic}.
Since the value of $\alpha$ is small, the value of $\alpha(s)$ given by 
Eq.\eqref{rg2} is practically indistinguishable from that given by \eqref{rg1}. The 
Landau singularity in Eq.\eqref{rg2} matters only when $s \sim e^{-\frac{2\pi}{\alpha}}$, so that 
\begin{equation}
\frac{2\pi}{\alpha} \approx R_1(s) = 1-\gamma-2\ln 2 - \ln s + \frac{3}{2}s + O(s^2),
\end{equation}
with $\gamma\approx0.5772$ being the Euler-Mascheroni constant.

\section{Conclusions \label{concl:sec}}
The basic symmetries of quantum electrodynamics are the relativistic invariance and the gauge invariance. In the standard 
approach to QED, which assumes the quantum fields to be local square-integrable functions, the calculations 
of observable quantities may violate both the Lorentz symmetry and the gauge symmetry due to the formal infinities 
of the calculated Green functions. Different regularization schemes have been used to get rid of divergences.
%%%%%%%%%%%%% Rev Nov 1 %%%%%%%%%%%%%%% 
The momentum cutoff regularization was historically the first. In low-energy {\em effective} theories there 
is a natural cutoff momentum $\Lambda$, above which the theory loses its validity. However, this cannot be 
applied considering a {\em fundamental} theory rather than effective theory.  
%%%%%%%%%%%%%%%%%%%%%%%%%%%%%%%%%%%%%%%%%%%%%%%%%%%%%%%%%%%%%%%%%%%%%%%%%%%%%%%%
The dimensional regularization has become the most common, utmost a standard way of regularization, because it does not 
violate the gauge symmetry; although it is not ubiquitous being incapable of treating supersymmetric theories
\cite{Stockinger2005}. 

Wavelet regularization is different from all  above mentioned regularizations. It changes the space of functions from 
the space of square-integrable functions to the space of functions $\psi_a(x)$, depending on both the coordinate $x$, 
and the scale $a$. The former is dynamical -- it enters the dynamical equations, the latter describes only the 
settings of observations and does not enter any dynamical equations. In this sense, we extend the description 
of observed physical fields by incorporating the conditions of observation ($a,\chi$) in the field definition.
A physical field {\sl per se} is then a collection of all physical fields that can be potentially observed: 
$\Psi = \{\psi_a(x,\cdot)\}_{a,\chi,\ldots}$. It cannot keep the perturbation expansion locally gauge invariant. Since the scale-dependent fields are defined not in a sharp point 
$x$, but in a region of typical size $a$, there is no need of infinite momentum injection to measure such fields,
and there are no physical reasons for the appearing of UV divergences. 

%%%%%%%%%%% Rev Oct 29
The key issue of the theory of scale-dependent 
fields is the problem of how the physical particles interact with each other. The description of physical interactions 
is determined by the symmetry of the problem. In this way, the symmetry with respect to local $U(1)$ transformations
determines electromagnetic interaction, the symmetry with respect to $SU(3)$ gauge transformations determines 
the strong interactions, and so on. In the present paper, we have followed exactly the same way: electrodynamics is understood as a theory with $U(1)$ gauge group, {\em acting on the space of square-integrable local functions}.
This definition immediately implies that the physically observed fields $\psi_a(x)$ are merely  projections 
of square-integrable fields $\psi(x)$ performed with the help of a mother wavelet $\chi$. This is a rather 
strong restriction: it states that gauge interactions take place in the space of local square-integrable functions and inverse wavelet transform is used to reconstruct the local fields from a set of their projections; the interaction
then takes place between the reconstructed fields. In this sense, wavelet regularization is similar to momentum cutoff 
regularization and gives the dependence of physical parameters on the observation scale.

Having performed the calculations, presented in this paper, we have found out that wavelet regularization 
can give a qualitatively adequate description of the QED running coupling constant (in one-loop approximation), 
which increases with the logarithm of the inverse observation scale. The advantage of wavelet regularization, 
if compared to dimensional regularization and other methods, is that it does not need any {\em external} tools, 
such as renormalization, which is always demanded by dimensional regularization to get physically interpretable 
results. The reason is that the wavelet transform itself is already based on the group of scale transformations, 
similar to the renormalization group. That is why instead of the renormalization group equations we have just a logarithmic 
derivative of the effective charge with respect to the dimensionless scale argument $s$. This $s=(\cA p)^2$ is similar to the normalization 
scale $1/\mu^2$ in dimensional regularization but has a physical interpretation in terms of the measurement scale.
The crucial difference from standard renormalization group approach is that we face no divergences to get rid of 
and we have no field renormalization. The latter is due to the fact that by extending the space of fields 
$\psi(x)$ to the space of scale-dependent fields $\psi_a(x)$ we already get the collections of {\em all} scales
rather than a poor man collection of two scales only. 
At the same time, our calculations show that known results such as Landau pole of the form $\frac{1}{1-\alpha X}$ 
also take place in a wavelet theory finite by construction, but in the latter case they have a more mild  
form of $1+\alpha X$, where $\alpha X$ is small and there is no threat of a pole.

At the same time, we have to admit that our persistence on keeping the standard definition of gauge invariance in the space of local field does not allow to preserve the transversality of the vacuum polarization operator at the
one-loop level: $p_\mu \Pi_{\mu\nu}(p)\ne0$. This has long been known for the momentum-cutoff and other 
regularization schemes and is quite expected for the wavelet regularization of a locally defined gauge 
theory: having declared the scale-dependent fields to be the physically observed fields we still insist that 
gauge interaction acts on local fields. It might be more reasonable do define the interaction directly in the 
space of scale-dependent fields, as is proposed in \cite{Altaisky2020PRD} in the context of QCD, but this is planned 
for future research.  
     
\section*{Acknowledgement}
The authors are thankful to Profs. M.Hnatich, S.Mikhailov and S.Thomas for useful discussions. The authors 
are also indebted to the anonymous referees for a series of useful comments. 
%\bibliography{qft}
\appendix
\newpage
\begin{widetext}
\section{Calculations with $\chi_1$ wavelet}
\subsection{Vacuum polarization diagram}
Substituting the integration measure $d^4q = 4\pi q^3 dq \sin^2\theta d\theta$ into the integral \eqref{padef} and dividing both the numerator and the denominator by $p^2q^2$ we arrive at the equation \eqref{Pag1}:
$$
\Pi_{\mu\nu}^{(\cA,\chi_1)} = - \frac{e^2 p^2}{\pi^3} e^{-\cA^2p^2}\int_0^\infty 
dy y e^{-4\cA^2p^2y^2} \int_0^\pi d\theta \sin^{2}\theta 
\frac{2y_\mu y_\nu - \frac{1}{2} \frac{p_\mu p_\nu}{p^2} + 
\delta_{\mu\nu}(\frac{1}{4}-y^2-\frac{m^2}{p^2})}
{\left[\frac{\frac{1}{4}+y^2+\frac{m^2}{p^2}}{y}+\cos\theta\right]
\left[\frac{\frac{1}{4}+y^2+\frac{m^2}{p^2}}{y}-\cos\theta\right]
}. \eqno(\ref{Pag1})
$$

The angle part of  integral \eqref{Pag1} can be evaluated explicitly. The corresponding integrals have the form:
\begin{align}
I_k[\beta(y)] = \int_0^\pi d\theta \frac{\sin^2\theta \cos^{2k}\theta}{\beta^2 - \cos^2\theta},  \quad 
I_0[\beta(y)] = \pi (1- \sqrt{1-\beta^{-2}}),\label{I0y}  \quad 
I_1[\beta(y)] = -\frac{\pi}{2} + \beta^2 I_0[\beta], 
% I_2[\beta(y)] &=& -\frac{\pi}{8}+\beta^2 \left[-\frac{\pi}{2} + \beta^2 I_0[\beta] \right]
\end{align}
where in our case  $\beta$ depends on $y$. To make the calculations analytically feasible, we simplify the matter by considering the relativistic limit 
$p^2 \gg 4m^2$. In this limit 
\begin{equation}
\beta(y) =  y + \frac{1}{4y}. \label{BT}
\end{equation}
In this approximation the vacuum polarization integral takes the form 
\begin{align*}
\Pi_{\mu\nu}^{(\cA,\chi_1)} = - \frac{e^2 p^2}{\pi^3} e^{-\cA^2p^2}\int_0^\infty 
dy y e^{-4\cA^2p^2y^2} 
\int_0^\pi d\theta \sin^{2}\theta 
\frac{2y_\mu y_\nu - \frac{1}{2} \frac{p_\mu p_\nu}{p^2} + 
\delta_{\mu\nu}(\frac{1}{4}-y^2)}
{\beta^2(y)-\cos^2\theta
}.
\end{align*}
There are two basic integrals here: the scalar integrals, which do not contain $y_\mu 
y_\nu$, and the tensor integrals, which contain these terms. The scalar integral has the form:
\begin{equation}
J \equiv \int_0^\infty dy y e^{-4sy^2} I_0[\beta(y)].
\end{equation}
As it follows from Eqs.(\ref{I0y},\ref{BT}):
\begin{equation}
I_0[\beta(y)] = \begin{cases}
 \frac{8\pi y^2}{1+4y^2}, & 0 \le y \le \frac{1}{2} \cr
\frac{2\pi}{1+4y^2}, & y \ge \frac{1}{2} 
\end{cases}, \label{I0py}\qquad
I_1[\beta(y)] = \begin{cases}
2\pi y^2, & 0 \le y \le \frac{1}{2} \cr
\frac{\pi}{8y^2}, & y \ge \frac{1}{2} 
\end{cases},
\end{equation}
and hence
\begin{align}
J = 8\pi \int_0^\frac{1}{2} \frac{dy y^3 e^{-4sy^2}}{1+4y^2}
+ 2\pi \int_{\frac{1}{2}}^\infty \frac{dy y e^{-4sy^2}}{1+4y^2} 
= \frac{\pi}{4}\left[
2 \Ei_1(2s) e^s - \Ei_1(s) e^s - \frac{e^{-s}}{s} + \frac{1}{s}.
\right]. \label{J-int}
\end{align}
The other scalar integral is a derivative of $J$:
\begin{align*}\nonumber 
I_q^0 = \int_0^\infty dy y^3 I_0[\beta(y)] e^{-4sy^2} = - \frac{1}{4}\frac{dJ}{ds} 
= - \frac{\pi}{16} \left[
2 \Ei_1(2s) e^s - \Ei_1(s) e^s - \frac{e^{-s}-1}{s} + \frac{e^{-s}-1}{s^2} 
\right].
\end{align*}
The terms of the integral, which contain $y_\mu y_\nu$, and which do not, can be evaluated separately. 
Since our problem has only one preferable direction -- the direction 
of vector $\mathbf{p}$, we can evaluate the tensor integral 
\begin{align}
I^{(2)}_{q\mu\nu} = \int y dy \sin^2 \theta d\theta e^{-4sy^2}
\frac{y_\mu y_\nu}
{\left(y + \frac{1}{4y} \right)^2 - \cos^2\theta}
\end{align}
by substituting 
$ y_\mu y_\nu \to B y^2 \delta_{\mu\nu} + C y^2 \frac{p_\mu p_\nu}{p^2}$ in the integrand; where $B$ and $C$ are the constants to be determined. 

The integral to be evaluated takes the form 
\begin{equation}
I_{\mu\nu} = \int y dy \sin^2 \theta d\theta e^{-4sy^2}
\frac{B y^2 \delta_{\mu\nu} + C y^2 \frac{p_\mu p_\nu}{p^2}}
{\left(y + \frac{1}{4y} \right)^2 - \cos^2\theta}.
\end{equation}

The constants $B$ and $C$ are determined from a system of two 
linear equation 
\begin{align*}
\Tr I^{(2)}_{q\mu\nu} = \Tr I_{\mu\nu}, \quad
I^{(2)}_{q\mu\nu} p_\mu p_\nu = I_{\mu\nu}p_\mu p_\nu\qquad \Leftrightarrow 4B+C=1, \quad
I_q^1 = (B+C)I_q^0,
\end{align*}
%or equivalently 
%\begin{align}
%4B+C=1, \quad
%I_q^1 = (B+C)I_q^0,
%\end{align}
where 
\begin{align}\nonumber 
I_q^1 = \int_0^\infty y^3 dy e^{-4sy^2} I_1[\beta(y)]=
\frac{\pi}{32}\frac{e^{-s}}{s^3}\left[e^s-s-1 \right].
\label{Iq}
\end{align}
Hence we can determine the constants $B$ and $C$ in terms of $\eta \equiv \frac{I_q^1}{I_q^0}$:
\begin{align}\nonumber 
B = \frac{1-\eta}{3}, \quad C = \frac{4\eta-1}{3}, \qquad
\eta = \frac{1}{2} \frac{1+s-e^{-s}}{
s^3 e^{2s}(2\Ei_1(2s)-\Ei_1(s))+s(s-1)(e^s-1)},
\end{align}
and now can evaluate the vacuum polarization diagram
\begin{align}
\Pi_{\mu\nu} &= - \frac{e^2p^2}{\pi^3}e^{-s}\left(I_q + I_p + I_t \right) 
\equiv I_T \delta_{\mu\nu} + I_P \frac{p_\mu p_\nu}{p^2}, \\
\nonumber
I_t &= \frac{1}{2} \left(\delta_{\mu\nu}-\frac{p_\mu p_\nu}{p^2} \right) J, \qquad 
I_p = - \frac{\delta_{\mu\nu}}{4}J, \quad 
I_q = 2 I^{(2)}_{q\mu\nu}-\delta_{\mu\nu}I^0_q = \left[\delta_{\mu\nu}(2B-1) 
+2C \frac{p_\mu p_\nu}{p^2} \right]I^0_q, \\
I_T &= -\frac{e^2p^2}{\pi^3}e^{-s}\left[ (2B-1)I_q^0 + \frac{J}{4} \right], \quad 
I_P = -\frac{e^2p^2}{\pi^3}e^{-s}\left[ 2CI_q^0 - \frac{J}{2} \right]. \label{X}
\end{align}
After simple algebra, we get the final expressions: 
\begin{align}
I_T &= \frac{e^2p^2}{48\pi^2s^3}
\Bigl[
(4s^2-2s-1)e^{-2s} +(1+s-4s^2)e^{-s} 
+ 4s^3 (\Ei_1(s)-2\Ei_1(2s))\Bigr] \label{ID} \\
I_P &= \frac{e^2p^2}{48\pi^2s^3}
\Bigl[
(-4s^2+2s+4)e^{-2s} +2(2s^2+s-2)e^{-s} 
- 4s^3 (\Ei_1(s)-2\Ei_1(2s))\Bigr], \label{IP}\\
\nonumber 
I_L &= I_T+I_P = \frac{e^2p^2}{16\pi^2 s^3} e^{-2s}\left((s-1)e^s+1\right).
\end{align}
We can represent the vacuum polarization diagram is a standard form:
\begin{align*}
\Pi_{\mu\nu} &= I_T\left(\delta_{\mu\nu} - \frac{p_\mu p_\nu}{p^2}\right) 
+ I_L \frac{p_\mu p_\nu}{p^2}.
\end{align*}
%\end{widetext}

\subsection{One-loop corrections to the QED vertex calculated with $\chi_1$ wavelet cutoff}
%\begin{widetext}
To evaluate the  integral \eqref{loop1} we take the tensor structure in the form 
$$
A_\rho = 2 \slashed{l}\gamma_\rho\slashed{l}- \frac{1}{2}\slashed{p}\gamma_\rho\slashed{p},
$$
omitting the terms linear in $l$ and using the massless limit.
Thus the whole integral in \eqref{loop1} takes the form:
\begin{align} 
I_\triangle = \int \frac{d^4y}{y^4p^2(2\pi)^4}\frac{
e^{-6\cA^2p^2y^2}\left(2p^2\slashed{y}\gamma_\rho \slashed{y} - \frac{1}{2}\slashed{p}\gamma_\rho \slashed{p}\right) 
}{
\left(y+\frac{1}{4y} \right)^2-\cos^2\theta
}  
= 2 \gamma_\alpha \gamma_\rho \gamma_\beta I^\triangle_{\alpha\beta} 
- \frac{\slashed{p}\gamma_\rho\slashed{p}}{2p^2} I_C,  \label{Itriangle}\\
I^\triangle_{\alpha\beta} = \frac{1}{4\pi^3} \int  \frac{d\theta \sin^2\theta dy y
e^{-6\cA^2p^2y^2}   
}{
\left(y+\frac{1}{4y} \right)^2-\cos^2\theta
} 
\frac{y_\alpha y_\beta}{y^2}, \quad   \label{It1}
I_C = \frac{1}{4\pi^3} \int  \frac{d\theta \sin^2\theta dy y
e^{-6\cA^2p^2y^2}   
}{
\left(y+\frac{1}{4y} \right)^2-\cos^2\theta
} \frac{1}{y^2},
\end{align} 
where we have used dimensionless momentum: $l = |p|y$, and have changed to the integration in spherical coordinates.

First, we evaluate the scalar integral $I_C$. Integrating over the angle variable $\theta$, we get: 
$$I_C = \frac{1}{4\pi^3} \int_0^\infty \frac{y dy}{y^2} I_0[\beta(y)] e^{-6\cA^2p^2}.$$
Since $I_0[\beta(y)]$ is a piecewise defined function \eqref{I0py}, we get 
\begin{align}\nonumber 
I_C = \frac{1}{4\pi^2} \int_0^1 \frac{dt e^{-\frac{3}{2}st}}{1+t} 
+\frac{1}{4\pi^2} \int_1^\infty \frac{dt e^{-\frac{3}{2}st}}{t(1+t)} 
= \frac{1}{4\pi^2} \left[
\Ei_1\left(\frac{3s}{2}\right) \left(1+e^{\frac{3s}{2}} \right) -2\Ei_1(3s)e^{\frac{3s}{2}}
\right],
\end{align}
where we have changed to a new variable $t=4y^2$ and used dimensionless scale argument $s=\cA^2p^2$.

Next, we evaluate the tensor integral. Since we have only one preferable direction -- that of $\vec{p}$ 
we can find the integral $I^\triangle_{\alpha\beta}$ in the form: 
\begin{equation}
\int  \frac{d\theta \sin^2\theta dy y
e^{-6\cA^2p^2y^2}   
}{
\left(y+\frac{1}{4y} \right)^2-\cos^2\theta
} 
\frac{y_\alpha y_\beta}{y^2} = \left(B \delta_{\alpha\beta} + C \frac{p_\alpha p_\beta}{p^2}\right) \int  \frac{d\theta \sin^2\theta dy y
e^{-6\cA^2p^2y^2}   
}{
\left(y+\frac{1}{4y} \right)^2-\cos^2\theta
} \label{BC3}
\end{equation}
where $B$ and $C$ are unknown constants to be determined. Tracing of both sides of equality \eqref{BC3} 
gives the constraint $4B+C=1$. Taking the convolution of both sides of \eqref{BC3} with $\frac{p_\mu p_\nu}{p^2}$,
we get another constraint
\begin{equation}
B+C = I_1/I_0, \quad I_1 = \int  \frac{d\theta\cos^2\theta \sin^2\theta dy y
e^{-6\cA^2p^2y^2}   
}{
\left(y+\frac{1}{4y} \right)^2-\cos^2\theta
}, 
\quad I_0 = \int  \frac{d\theta \sin^2\theta dy y
e^{-6\cA^2p^2y^2}   
}{
\left(y+\frac{1}{4y} \right)^2-\cos^2\theta
}. \label{IBC}
\end{equation}
The integral $I_0$ in Eq.\eqref{IBC} coincides with the integral $J$, given by \eqref{J-int} up to the 
change of scale $s \to \frac{3}{2}s$. This gives:
\begin{equation}
I_0 = \frac{\pi}{4}\left[
2\Ei_1(3s)e^\frac{3s}{2} - \Ei_1\left(\frac{3s}{2} \right) e^\frac{3s}{2} 
- \frac{2e^{-\frac{3s}{2}}}{3s} + \frac{2}{3s}
\right].
\end{equation}
After the angle integration in $I_1$, we get $I_1=\int_0^\infty y dy e^{-6sy^2}I_1[\beta(y)]$, with $I_1[\beta(y)]$ 
given by Eq. \eqref{I0py}, from where we get:
\begin{equation}
I_1 = \frac{\pi}{16} \int_0^1 dt t e^{-\frac{3}{2}st} + \frac{\pi}{16} \int_1^\infty dt \frac{ e^{-\frac{3}{2}st}}{t} =\frac{\pi}{16} \left[
\Ei_1\left(\frac{3s}{2} \right) + \frac{4}{9s^2} - \frac{4 e^{-\frac{3s}{2}}}{9s^2} - \frac{2e^{-\frac{3s}{2}} }{3s}
\right],
\end{equation}
where  $t=4y^2$.

We can rewrite \eqref{Itriangle} in the form 
\begin{equation}
I_\triangle = 2 \gamma_\alpha \gamma_\rho \gamma_\beta \frac{I_0}{4\pi^3}\left(
B \delta_{\alpha\beta} + C \frac{p_\alpha p_\beta}{p^2}\right) - \frac{\slashed{p}\gamma_\rho \slashed{p}}{2p^2}I_C=
\left(\frac{I_0}{2\pi^3}(2B+C) - \frac{I_C}{2} \right)\gamma_\rho 
+ \frac{\slashed{p}p_\rho}{p^2}\left(I_C - \frac{CI_0}{\pi^3} \right), \label{itbc}
\end{equation}
where the last term is not proportional to $\gamma_\rho$ and will be ignored.
Now we can substitute the found constants: 
$$ B = \frac{1-\eta}{3}, \quad C = \frac{4\eta-1}{3}, \quad \eta \equiv \frac{I_1}{I_0} $$
into equation \eqref{itbc} to obtain the one loop contribution to the fermion-photon vertex:
\begin{align} \label{lambda1rho}
\Lambda_\rho(-p/2,-p/2,p) &= e_0^2 e^{-s} I_\triangle = e^2_0 e^{-s} \left[ \frac{I_0}{6\pi^3}+  \frac{I_1}{3\pi^3} - \frac{I_C}{2}\right] \gamma_\rho \\  \nonumber
&= \frac{e_0^2 \gamma_\rho}{3\pi^2}\left[ 
e^\frac{s}{2}\Ei_1(3s) - \frac{e^\frac{s}{2}\Ei_1\left(\frac{3s}{2} \right)}{2} 
- \frac{e^{-\frac{5s}{2}}}{8s} + \frac{e^{-s}}{12s} 
-\frac{ 5 e^{-s}\Ei_1\left(\frac{3s}{2} \right)}{16}  + \frac{e^{-s}}{36s^2}
- \frac{e^{-\frac{5s}{2}}}{36s^2} 
 \right].
\end{align}

\section{Calculations with $\chi_2$ wavelet}
\subsection{Electron self-energy diagram}
Similarly to the case of the $\chi_1$ wavelet, we make the one-loop calculation 
with the $\chi_2$ wavelet by symmetrizing the loop momenta ($\frac{p}{2}+q,\frac{p}{2}-q$) 
for both the vacuum polarization and the electron self-energy diagrams. Thus, the 
wavelet cutoff factor in this diagrams is: 
$$F_\cA(p,q)=f^2\left(\cA\bigl( \frac{p}{2}+q\bigr) \right) 
f^2\left(\cA\bigl( \frac{p}{2}+q\bigr) \right),$$
with $f(x)$ for $\chi_2$ given by \eqref{fch1}. 
\begin{equation} 
F_\cA(p,q)=\cA^8 \left[\left(q^2+\frac{p^2}{4}+\frac{1}{\cA^2}\right)^2-p^2q^2\cos^2\theta\right]^2
e^{-\cA^2p^2-4\cA^2q^2}
= s^4 e^{-s}\left[\left(y+\frac{1}{4y}+\frac{1}{sy}\right)^2-\cos^2\theta\right]^2e^{-4sy^2}y^4
\label{fch2s}
\end{equation}
%where we have used dimensionless scale  $s=A^2p^2$. 
Now we can substitute $F_\cA(p,|p|y)$ 
into the equation for the electron self-energy \eqref{ese-y}. We omit the term $4m -2|p|\slashed{y}$ in the numerator, since $m$ is small in comparison to $\slashed{p}$ in our approximation, and $\slashed{y}$ does not contribute for symmetry reasons. This 
gives:
\begin{align}\nonumber 
\Sigma^{(\cA)}(p)=-\imath e^2 \slashed{p} s^4 e^{-s} \int \frac{d^4y}{(2\pi)^4} y^2 e^{-4sy^2} 
\frac{\left[ \left(y+\frac{1}{4y} +\frac{1}{sy}\right)^2-\cos^2\theta\right]^2}{\left(y+\frac{1}{4y} \right)^2-\cos^2\theta}.
\end{align}
Using the notation $\beta \equiv y+\frac{1}{4y}$ and the integration measure 
$d^4y =4\pi \sin^2\theta d\theta y^3 dy$, we get
\begin{align*}
\Sigma^{(\cA)}(p)&= -\imath \frac{e^2 e^{-s}\slashed{p}}{4\pi^3} \int \Bigl[ 
\left(\beta^2-\cos^2\theta \right)^2 s^4 y^4 + 
4 \beta \left(\beta^2-\cos^2\theta  \right) s^3 y^3 
+ 2 \left(3\beta^2-\cos^2\theta  \right) s^2 y^2 \\
&+ 4\beta s y + 1\Bigr] \frac{\sin^2\theta d\theta dy y e^{-4sy^2}}{\beta^2-\cos^2\theta} 
\end{align*}
We perform the angle integration first:
\begin{align}\nonumber 
\Sigma^{(\cA)}(p)= -\imath \frac{e^2 e^{-s}\slashed{p}}{4\pi^3} \int_0^\infty
dy y e^{-4sy^2} \int_0^\pi d\theta \sin^2\theta  
\Bigl[\bigl( \beta^2- \cos^2\theta\bigr) s^4 y^4 +4\beta s^3 y^3 + \\ \nonumber 
+ 2s^2y^2 \left(\frac{2\beta^2}{\beta^2-\cos^2\theta} +1\right)
+ \frac{4\beta s y +1}{\beta^2- \cos^2\theta} 
\Bigr] \\  
= -\imath \frac{e^2 e^{-s}\slashed{p}}{4\pi^3} \int_0^\infty dy y e^{-4sy^2}\Bigl\{
\frac{\pi}{2}\Bigl[ \bigl(\beta^2-\frac{1}{4}\bigr) s^4 y^4 
+ 4\beta s^3 y^3 + 2s^2y^2
\Bigr] + I_0[\beta]\bigl(4s^2y^2\beta^2+4\beta s y + 1 \bigr) \Bigr\}. \label{eseJ}
\end{align}
Since the angle integral $I_0[\beta(y)]$, given by Eq.\eqref{I0y}, is defined 
piecewise  \eqref{I0py}, we split $\Sigma^{(A)}(p)$ into a sum of two integrals: 
$$
\Sigma^{(A)}(p)\equiv J_1 + J_2,
$$
with the second one, which depends on $I_0[\beta(y)]$, to be splited as:$\int_0^\infty= \int_0^{1/2} + \int_{1/2}^\infty$. The integrals are:
%\begin{widetext}
\begin{align*}\nonumber
J_1 &= -\imath \frac{e^2e^{-s}\slashed{p}}{32\pi^2}\int_0^\infty 
\Bigl[ s^4y^4 \bigl(4y^2+\frac{1}{4y^2}+1\bigr) 
+ 16s^3y^3 \bigl(y+\frac{1}{4y} \bigr)
+8s^2y^2\Bigr] e^{-4sy^2}y dy   
=-\imath \frac{e^2e^{-s}\slashed{p}}{32\pi^2} 
\frac{s^2+18s+70}{128}
, \\ \nonumber
J_2 &= -\imath \frac{e^2e^{-s}\slashed{p}}{4\pi^3}\int_0^\infty I_0[\beta(y)]
\bigl(4s^2\beta^2y^2 
+4\beta s y +1 \bigr)e^{-4sy^2}y dy \\
& = -\imath \frac{e^2e^{-s}\slashed{p}}{4\pi^2}\Bigl\{
\int_0^{1/2} \frac{8y^2}{1+4y^2}\bigl(4s^2y^2\beta^2+4\beta s y+1 \bigr)e^{-4sy^2}y dy
+ \int_{1/2}^\infty \frac{2}{1+4y^2}\bigl(4s^2y^2\beta^2+4\beta s y+1 \bigr)e^{-4sy^2}y dy
\Bigr\} \\
&= -\imath \frac{e^2e^{-s}\slashed{p}}{64\pi^2}\Bigl\{
\int_0^{1} \frac{t}{t+1} \bigl(s^2(t+1)^2 +4s(t+1)+4 \bigr)e^{-st}dt
+ \int_1^{\infty} \frac{1}{t+1} \bigl(s^2(t+1)^2 +4s(t+1)+4 \bigr)e^{-st}dt \\
&= -\imath \frac{e^2e^{-s}\slashed{p}}{64\pi^2} \Bigl(
4e^{s}(2\Ei_1(2s) - \Ei_1(s)) 
 + \frac{10}{s}(1-e^{-s}) +1 - 2e^{-s}
\Bigr)
\end{align*}
The final result is: 
\begin{align} 
\Sigma^{(\cA)}_{\chi_2}(p) = -\frac{\imath e^2\slashed{p}}{16\pi^2}\Bigl[
 2\Ei_1(2s)  - \Ei_1(s)
-\frac{s+5}{2s} e^{-2s}
+ \frac{s^3+18s^2+134s+640}{256s}e^{-s}\Bigr].
 \label{sigma-chi2}
\end{align}

\subsection{Vacuum polarization diagram}
Calculation of vacuum polarization diagram for the case of the $\chi_2$ wavelet cutoff function \eqref{fch1} is 
completely analogous to that performed with $\chi_1$ wavelet cutoff. To simplify analytical calculation here we 
also assume relativistic limit $p^2 \gg 4m^2$ and omit appropriate terms. In this way, Eq. \eqref{padef} 
becomes: 
\begin{align*}\nonumber 
\Pi_{\mu\nu}^{(\cA),\chi_2}(p) = - \frac{e^2p^2}{\pi^3} \int_0^\infty dy y \int_0^\pi d\theta \sin^2\theta F_\cA(p,q)
 \frac{
2y_\mu y_\nu - \delta_{\mu\nu}y^2 +\frac{1}{2}\left(\delta_{\mu\nu} - \frac{p_\mu p_\nu}{p^2} \right)-\frac{\delta_{\mu\nu}}{4}
}{\beta^2(y)-\cos^2\theta}.
\end{align*} 
The wavelet cutoff function $F_\cA(p,q)$ is given by Eq.\eqref{fch2s}. In dimensionless variables $(s,y)$ it 
has the form: 
$$
F_\cA(y) =  e^{-s}e^{-4sy^2} s^4y^4 \left[ \left(\beta(y)+\frac{1}{sy} \right)^2-\cos^2\theta
\right]^2.
$$
So, the integral to be evaluated is: 
\begin{align}\nonumber 
\Pi_{\mu\nu}^{(\cA),\chi_2}(p) &= - \frac{e^2p^2e^{-s}}{\pi^3} \int_0^\infty dy y e^{-4sy^2} \int_0^\pi d\theta \sin^2\theta 
s^4 y^4 \left[ \left(\beta(y)+\frac{1}{sy} \right)^2-\cos^2\theta
\right]^2 \times \\ \nonumber 
&\times \frac{
2y_\mu y_\nu - \delta_{\mu\nu}y^2 +\frac{1}{2}\left(\delta_{\mu\nu} - \frac{p_\mu p_\nu}{p^2} \right)-\frac{\delta_{\mu\nu}}{4}
}{\beta^2(y)-\cos^2\theta}.
\end{align}
%\end{widetext}
Similar to the evaluation of the self-energy diagram, we expand the polynomial part of the wavelet cutoff 
function and get:
%\begin{widetext}
\begin{align}
\Pi_{\mu\nu}^{(A),\chi_2}(p) &= - \frac{e^2p^2e^{-s}}{\pi^3} \int_0^\infty dy y e^{-4sy^2} \int_0^\pi d\theta \sin^2\theta 
\Bigl[ s^4y^4 (\beta^2-\cos^2\theta) + 4\beta s^3y^3 + \label{po2full}\\
\nonumber &+2s^2 y^2 \bigl(\frac{2\beta^2}{\beta^2-\cos^2\theta}+1 \bigr) 
+\frac{4\beta s y+1}{\beta^2-\cos^2\theta}\Bigr] \times 
\Bigl[
2y_\mu y_\nu - \delta_{\mu\nu}y^2 +\frac{1}{2}\bigl(\delta_{\mu\nu} - \frac{p_\mu p_\nu}{p^2} \bigr)-\frac{\delta_{\mu\nu}}{4}
\Bigr].
\end{align}
To calculate the vacuum polarization diagram, Eq.\eqref{po2full}, we need two integrals. The scalar 
integral, and the tensor integral dependent on $y_\mu y_\nu$. They are: 
\begin{align}
J &= \int_0^\infty dy y e^{-4sy^2} \int_0^\pi d\theta \sin^2\theta 
\Bigl[ s^4y^4 (\beta^2-\cos^2\theta) + 4\beta s^3y^3 
+2s^2 y^2 \bigl(\frac{2\beta^2}{\beta^2-\cos^2\theta}+1 \bigr) 
+\frac{4\beta s y+1}{\beta^2-\cos^2\theta}\Bigr], \\
I^{(2)}_{q\mu\nu} &= \int_0^\infty dy y e^{-4sy^2} \int_0^\pi d\theta \sin^2\theta 
\Bigl[ s^4y^4 (\beta^2-\cos^2\theta) + 4\beta s^3y^3 
+2s^2 y^2 \bigl(\frac{2\beta^2}{\beta^2-\cos^2\theta}+1 \bigr) 
+\frac{4\beta s y+1}{\beta^2-\cos^2\theta}\Bigr]y_\mu y_\nu. \label{I2qmn}
\end{align}
The integral $J$ is identical to that calculated for electron self-energy diagram in Eq.~\eqref{eseJ}.
Its value is: 
\begin{equation}
J = \frac{\pi}{4}\Bigl[
\frac{s^3+18s^2+134s+640}{256s} - \frac{e^{-s}(s+5)}{2s} + e^s \bigl(2\Ei_1(2s)-\Ei_1(s) \bigr)
\Bigr].
\end{equation}
The integral \eqref{I2qmn} is evaluated by changing $y_\mu y_\nu \to B y^2 \delta_{\mu\nu} + C y^2 \frac{p_\mu\nu}{p^2}$:
\begin{align*}
I_{\mu\nu} &= \int_0^\infty dy y e^{-4sy^2} \int_0^\pi d\theta \sin^2\theta 
\Bigl[ s^4y^4 (\beta^2-\cos^2\theta) + 4\beta s^3y^3 + \\
 &+2s^2 y^2 \bigl(\frac{2\beta^2}{\beta^2-\cos^2\theta}+1 \bigr) 
+\frac{4\beta s y+1}{\beta^2-\cos^2\theta}\Bigr]\bigl[B y^2 \delta_{\mu\nu} + C y^2 \frac{p_\mu p_\nu}{p^2} \bigr].
\end{align*}
The unknown constants $B$ and $C$ are determined from the equality 
$I^{(2)}_{q\mu\nu} = I_{\mu\nu}$, exactly in the same way as for the $\chi_1$ wavelet. Taking the trace of both sides we get the constraint $4B+C=1$. The other 
constraint is obtained by convolution of both sides of $I^{(2)}_{q\mu\nu} = I_{\mu\nu}$ with $\frac{p_\mu p_\nu}{p^2}$. This gives 
$I^0_q (B+C) = I^1_q$, where 
\begin{align*}
I^0_q &= \int_0^\infty dy y^3 e^{-4sy^2} \int_0^\pi d\theta \sin^2\theta 
\Bigl[ s^4y^4 (\beta^2-\cos^2\theta) + 4\beta s^3y^3  
+2s^2 y^2 \bigl(\frac{2\beta^2}{\beta^2-\cos^2\theta}+1 \bigr) 
+\frac{4\beta s y+1}{\beta^2-\cos^2\theta}\Bigr]  \\
&= \int_0^\infty dy y^3 e^{-4sy^2} \Bigl[ 
\frac{\pi}{2} \bigl(\beta^2 s^4 y^4 + 4\beta s^3 y^3 + 2s^2y^2 \bigr) - \frac{\pi}{8} s^4 y^4 
+ I_0[\beta(y)] \bigl(4\beta^2 s^2 y^2 + 4\beta s y + 1 \bigr),\\
I^1_q &= \int_0^\infty dy y^3 e^{-4sy^2} \int_0^\pi d\theta \sin^2\theta \cos^2\theta
\Bigl[ s^4y^4 (\beta^2-\cos^2\theta) + 4\beta s^3y^3  
+2s^2 y^2 \bigl(\frac{2\beta^2}{\beta^2-\cos^2\theta}+1 \bigr) 
+\frac{4\beta s y+1}{\beta^2-\cos^2\theta}\Bigr]  \\
&= \int_0^\infty dy y^3 e^{-4sy^2} \Bigr[ \frac{\pi}{8} \bigl( s^4 y^4 \beta^2 +4\beta s^3 y^3+ 2s^2 y^2)  -\frac{\pi}{16}s^4y^4 
+I_1[\beta(y)] \bigl(4\beta^2 s^2 y^2 + 4\beta s y + 1 \bigr) \Bigr]. 
\end{align*}
The part of the integral, which depends on piecewise-defined functions $I_0[\beta(y)],I_1[\beta(y)]$, given by Eq.(\ref{I0py}),
is integrated in $\int_0^{1/2}+\int_{1/2}^\infty$ limits, accordingly. This gives:
\begin{align}
I^0_q &= \frac{9\pi(1-e^{-s})}{32s^2} + \frac{7\pi}{512s}-\frac{\pi e^{-s}}{32} - \frac{3\pi e^{-s}}{32s}  + \frac{19\pi}{2048} + \frac{\pi s}{2048} - \frac{\pi e^s}{16}(2\Ei_1(2s) - \Ei_1(s)), \\
I^1_q &= \frac{7\pi(1-e^{-s})}{32s^3} + \frac{5\pi}{64s^2} + \frac{39\pi}{2048s} 
-\frac{19\pi e^{-s}}{64s^2}
-\frac{5\pi e^{-s}}{32s}-\frac{\pi e^{-s}}{32} + \frac{\pi s}{8192} + \frac{\pi}{512} . 
\end{align}
Using the expression \eqref{X} we get the equation for the vacuum polarization diagram in the form:
$$\Pi_{\mu\nu}^{(\cA),\chi_2} = I_T \delta_{\mu\nu} + I_P \frac{p_\mu p_\nu}{p^2}, \quad 
I_T = - \frac{e^2 p^2 e^{-s}}{\pi^3} \left[
-\frac{1}{3}I^0_q-\frac{2}{3}I^1_q + \frac{J}{4} \right], \quad 
I_P = - \frac{e^2 p^2 e^{-s}}{\pi^3} \left[
\frac{2}{3}(4I^1_q-I^0_q) - \frac{J}{2}
\right].
$$
This gives
\begin{align}
\nonumber I_T &= \frac{e^2p^2}{\pi^2}\Bigg[
-\frac{1}{12}(2\Ei_1(2s)-\Ei_1(s)) \\ 
& -\frac{s^2 e^{-s}}{4096} - \frac{17s e^{-s}}{4096} - \frac{427e^{-s}}{48s} + \frac{7e^{-s}}{48s^2} + \frac{7e^{-s}}{48s^3}
- \frac{29e^{-s}}{1024} + \frac{e^{-2s}}{48s} - \frac{7 e^{-2s}}{24s^2} - \frac{7e^{-2s}}{48s^3} 
\Bigg], \\
\nonumber I_P &= \frac{e^2p^2}{\pi^2}\Bigg[
\frac{1}{12}(2\Ei_1(2s)-\Ei_1(s)) \\ 
& +\frac{s^2 e^{-s}}{2048} + \frac{9s e^{-s}}{1024} + \frac{13e^{-s}}{48s} - \frac{e^{-s}}{48s^2} - \frac{7e^{-s}}{12s^3}
+ \frac{17e^{-s}}{256} + \frac{e^{-2s}}{24s} + \frac{29 e^{-2s}}{48s^2} + \frac{7e^{-2s}}{12s^3} 
\Bigg].
\end{align}

\subsection{One-loop corrections to the QED vertex calculated with $\chi_2$ wavelet cutoff}
In complete analogy to the calculations performed with $\chi_1$ wavelet, we can express the 
wavelet cutoff function corresponding to the diagram shown Fig.~\ref{epv:pic} in the relativistic limit $p^2\gg 4m^2$, in the form 
\begin{equation}
F_\cA(p,l) = e^{-s} e^{-6sy^2} s^4 y^4 \left[ 
\left(y + \frac{1}{4y} + \frac{1}{sy} \right)^2 - \cos^2\theta
\right]^2 \left( 1+ sy^2 \right)^2, \quad \hbox{where\ } l = y|p|. \label{v32}
\end{equation}
The one-loop contribution to the fermion-photon vertex calculated with this cutoff is:
\begin{align}\nonumber 
\Lambda_\rho\left(-\frac{p}{2},-\frac{p}{2},p\right) &= -e_0^2 s^4 e^{-s} 
\int \frac{d^4y}{(2\pi)^4} \frac{
2 \slashed{y}\gamma_\rho \slashed{y} - \frac{\slashed{p}\gamma_\rho\slashed{p} }{2p^2}
}{
\left(y+\frac{1}{4y}\right)^2 - \cos^2\theta 
}
e^{-6sy^2}\left[ 
\left(y + \frac{1}{4y} + \frac{1}{sy} \right)^2 - \cos^2\theta
\right]^2 \left( 1+ sy^2 \right)^2 \\
& \equiv -e_0^2 e^{-s} s^4 I_\triangle. \label{lambda2}
\end{align}
The integral \eqref{lambda2} is a sum of two integrals:
$
I_\triangle = 2 \gamma_\alpha \gamma_\rho \gamma_\beta I_{\alpha\beta}^\triangle - \frac{\slashed{p}\gamma_\rho\slashed{p} }{2p^2} I_C
$, where 
\begin{align*}
I_C &= \int \frac{d^4y}{(2\pi)^4} \frac{
e^{-6sy^2}
}{
\left(y+\frac{1}{4y}\right)^2 - \cos^2\theta 
}
\left[ 
\left(y + \frac{1}{4y} + \frac{1}{sy} \right)^2 - \cos^2\theta
\right]^2 \left( 1+ sy^2 \right)^2, \\
I_{\alpha\beta}^\triangle &= \int \frac{d^4y}{(2\pi)^4} \frac{ y_\alpha y_\beta 
e^{-6sy^2}
}{
\left(y+\frac{1}{4y}\right)^2 - \cos^2\theta 
}
\left[ 
\left(y + \frac{1}{4y} + \frac{1}{sy} \right)^2 - \cos^2\theta
\right]^2 \left( 1+ sy^2 \right)^2.
\end{align*}
The evaluation of these integral is identical to the case of the $\chi_1$ wavelet, described by Eqs.~(\ref{It1},\ref{Itriangle}). Using the variable $\beta(y)=y+\frac{1}{4y}$, we get: 
\begin{align*} 
I_C &= \int \frac{d^4y}{(2\pi)^4} \frac{e^{-6sy^2}}{\beta^2-\cos^2\theta} \left(1+sy^2 \right)^2
\left[
\beta^2 -\cos^2\theta + \frac{2\beta}{sy} + \frac{1}{s^2y^2} 
\right]^2  \\ \nonumber 
&= \frac{1}{4\pi^3} \int dy d\theta y^3  \sin^2\theta  e^{-6sy^2}(1+sy^2)^2 (\beta^2-\cos^2\theta) 
+ \frac{1}{2\pi^3} \int dy d\theta y^3  \sin^2\theta  e^{-6sy^2}(1+sy^2)^2 
\left(\frac{2\beta}{sy} + \frac{1}{s^2y^2} \right) \\ \nonumber 
&+ \frac{1}{4\pi^3} \int \frac{dy d\theta y^3  \sin^2\theta  e^{-6sy^2}(1+sy^2)^2}{\beta^2 -\cos^2\theta}
\left(\frac{2\beta}{sy} + \frac{1}{s^2y^2} \right)^2 \\
&= \frac{1}{8\pi^2} \int_0^\infty dy y^3  e^{-6sy^2} (1+sy^2)^2 \left(\beta^2-\frac{1}{4} \right)
+ \frac{1}{4\pi^2} \int_0^\infty dy y^3  e^{-6sy^2} (1+sy^2)^2 \left(\frac{2\beta}{sy} + \frac{1}{s^2y^2} \right)
\\
&+\frac{1}{4\pi^3} \int_0^{1/2} dy y^3  e^{-6sy^2} (1+sy^2)^2 \left(\frac{2\beta}{sy} + \frac{1}{s^2y^2} \right)^2
\frac{8\pi y^2}{1+4y^2} 
+\frac{1}{4\pi^3} \int_{1/2}^\infty dy y^3  e^{-6sy^2} (1+sy^2)^2 \left(\frac{2\beta}{sy} + \frac{1}{s^2y^2} \right)^2
\frac{2\pi}{1+4y^2}, 
\end{align*}
where we have used the angle integration rule \eqref{I0py}. After the change of variable $t=4y^2$, the final 
result is: 
\begin{align} \nonumber 
I_C &= \frac{25}{27648\pi^2s} + \frac{211}{13824\pi^2s^2} + \frac{469}{5184 \pi^2 s^3} + \frac{161}{432\pi^2s^4} 
- \frac{e^{-\frac{3}{2}s}}{\pi^2 s^2} \left( 
\frac{1}{288} + \frac{7}{96s} + \frac{161}{432 s^2}
\right) \\
&+ \frac{e^{\frac{3}{2}s}}{4\pi^2 s^2} \left( 
\Ei_1 \left(\frac{3}{2}s\right) -2\Ei_1(3s) 
\right) \left( \frac{1}{16} - \frac{1}{2s} + \frac{1}{s^2}\right) 
+ \frac{\Ei_1 \left(\frac{3}{2}s \right)}{4\pi^2s^2} \left( 
\frac{1}{4}+ \frac{1}{s} + \frac{1}{s^2}\right).
\label{Ic2}
\end{align}
The tensor integral $I_{\alpha\beta}^\triangle$ can be decomposed with respect to two basic 
tensors $\delta_{\alpha\beta}$ and $\frac{p_\alpha p_\beta}{p^2}$:
\begin{align} \label{vt2}
I_{\alpha\beta}^\triangle &= \frac{1}{4\pi^3} \int \frac{dy d\theta \sin^2\theta  y^3  e^{-6sy^2} }{
\beta^2-\cos^2\theta}(1+sy^2)^2 \left[ \left(\beta + \frac{1}{sy} \right)^2-\cos^2\theta \right]^2 y_\alpha y_\beta 
\\ \nonumber 
&= \frac{1}{4\pi^3} \int \frac{dy d\theta \sin^2\theta y^3 e^{-6sy^2} }{
\beta^2-\cos^2\theta}(1+sy^2)^2 \left[ \left(\beta + \frac{1}{sy} \right)^2-\cos^2\theta \right]^2
\left[
B y^2 \delta_{\alpha\beta} + C y^2 \frac{p_\alpha p_\beta}{p^2}
\right],
\end{align}
where $B$ and $C$ are unknown constants to be determined.  Tracing of both sides of Eq.\eqref{vt2} gives the constraint 
$4B+C=1$. The other constraint is identical to \eqref{IBC}:
$$
I_1 = (B+C) I_0,
$$
where 
\begin{align}
I_0 &= \int \frac{dy d\theta \sin^2\theta  y^5 e^{-6sy^2} }{\beta^2-\cos^2\theta} 
(1+sy^2)^2 \left[ \left(\beta + \frac{1}{sy} \right)^2 -\cos^2\theta \right]^2, \\
I_1 &= \int \frac{dy d\theta \sin^2\theta \cos^2\theta y^5 e^{-6sy^2} }{\beta^2-\cos^2\theta} 
(1+sy^2)^2 \left[ \left(\beta + \frac{1}{sy} \right)^2 -\cos^2\theta \right]^2.
\end{align}
The remaining calculations are analogues to previous integrals:
\begin{align*}
I_0 &= \int dy d\theta \sin^2\theta y^5 e^{-6sy^2}(1+sy^2)^2 (\beta^2-\cos^2\theta) 
+ 2 \int dy d\theta \sin^2\theta y^5 e^{-6sy^2}(1+sy^2)^2 
\left( \frac{2\beta}{sy}+\frac{1}{s^2y^2} \right) \\
&+ \int \frac{
 dy d\theta \sin^2\theta y^5 e^{-6sy^2}(1+sy^2)^2
}{\beta^2-\cos^2\theta} \left( \frac{2\beta}{sy}+\frac{1}{s^2y^2} \right)^2 \\
&= \frac{\pi}{2}\int_0^\infty dy y^5 e^{-6sy^2}(1+sy^2)^2 \frac{16y^4+4y^2+1}{16y^2} 
+ \pi \int_0^\infty dy y^5 e^{-6sy^2}(1+sy^2)^2 \frac{4y^2 s + s +2}{2s^2y^2} \\
&+ \int_0^{1/2} dy y^5 \frac{8\pi y^2}{1+4y^2} e^{-6sy^2}(1+sy^2)^2 
\left( \frac{4y^2 s + s +2}{2s^2y^2}
\right)^2 
+ \int_{1/2}^\infty dy y^5 \frac{2\pi}{1+4y^2} e^{-6sy^2}(1+sy^2)^2 
\left( \frac{4y^2 s + s +2}{2s^2y^2}
\right)^2 \\
&= - \frac{\pi e^{-\frac{3s}{2}}}{124416 s^5}  \bigg( 14976s + 7056 s^2 - 64896 e^\frac{3s}{2} + 432s^3 -1376 s e^\frac{3s}{2} 
-3048 s^2 e^\frac{3s}{2} + 64896 -15552 e^{3s} s^2 \Ei_1(\frac{3s}{2}) \\
&+ 31104 e^{3s} s \Ei_1( \frac{3s}{2}) + 1944 e^{3s} s^3 \Ei_1( \frac{3s}{2}) -62208 s e^{3s} \Ei_1(3s) 
-3888 s^3 e^{3s} \Ei_1(3s) + 31104 s^2 e^{3s} \Ei_1(3s) -99 s^3 e^\frac{3s}{2}\bigg),\\
I_1 &= \int dy d\theta \sin^2\theta \cos^2\theta y^5 e^{-6sy^2}(1+sy^2)^2 (\beta^2-\cos^2\theta) 
+ 2 \int dy d\theta \sin^2\theta \cos^2\theta y^5 e^{-6sy^2}(1+sy^2)^2 
\left( \frac{2\beta}{sy}+\frac{1}{s^2y^2} \right) \\
&+ \int \frac{
 dy d\theta \sin^2\theta \cos^2\theta y^5 e^{-6sy^2}(1+sy^2)^2
}{\beta^2-\cos^2\theta} \left( \frac{2\beta}{sy}+\frac{1}{s^2y^2} \right)^2 \\
&= \frac{\pi}{8}\int_0^\infty dy y^5 e^{-6sy^2}(1+sy^2)^2 \frac{16y^4+1}{16y^2} 
+ \frac{\pi}{4} \int_0^\infty dy y^5 e^{-6sy^2}(1+sy^2)^2 \frac{4y^2 s + s +2}{2s^2y^2} \\
&+ \int_0^{1/2} dy y^5 2\pi y^2 e^{-6sy^2}(1+sy^2)^2 
\left( \frac{4y^2 s + s +2}{2s^2y^2}
\right)^2 
+ \int_{1/2}^\infty dy y^5 \frac{\pi}{8y^2} e^{-6sy^2}(1+sy^2)^2 
\left( \frac{4y^2 s + s +2}{2s^2y^2}
\right)^2 \\
&= \frac{\pi e^{-\frac{3s}{2}}}{497664 s^6}\bigg( 
99 s^4 e^\frac{3s}{2} + 12608 s^2 e^\frac{3s}{2} + 1584 s^3 e^\frac{3s}{2} + 46848s e^\frac{3s}{2} 
+ 94976 e^\frac{3s}{2} - 94976 -189312s -121536 s^2 -1728 s^4 \\
&- 28224s^3 + 7776s^4 e^\frac{3s}{2}\Ei_1(\frac{3s}{2}) + 31104 s^3 e^\frac{3s}{2}\Ei_1(\frac{3s}{2}) 
+ 31104s^2 e^\frac{3s}{2}\Ei_1(\frac{3s}{2})
\bigg),
\end{align*}
where we have used the angle integration rules \eqref{I0py}. Substituting these integrals into the 
final equation \eqref{lambda2}, and comparing it to \eqref{lambda1rho}, we get:
\begin{align} 
\Lambda_\rho^{\chi_2}\left(-\frac{p}{2},-\frac{p}{2},p\right) &= e_0^2 s^4 e^{-s} \left[
\frac{I_0}{6\pi^3} + \frac{I_1}{3\pi^3} - \frac{I_C}{2}
\right] \gamma_\rho = \\ \nonumber 
&= - \frac{\gamma_\rho e_0^2 e^{-\frac{5s}{2}}}{1492992\pi^2 s^2}\bigg( 59856 s^3 e^\frac{3s}{2} + 
319104s -126720 s^2  -12096 s^3 + 262848 s^2 e^\frac{3s}{2}- 176640 s e^\frac{3s}{2} \\ 
\nonumber &-94976 e^\frac{3s}{2} + 38880 s^4 e^\frac{3s}{2}\Ei_1(\frac{3s}{2})
+ 155520 s^3 e^\frac{3s}{2} \Ei_1(\frac{3s}{2}) 
+ 248832 e^{3s} s^2 \Ei_1( \frac{3s}{2}) \\
\nonumber &- 124416 e^{3s} s^3 \Ei_1( \frac{3s}{2}) + 248832 s^3 e^{3s}\Ei_1(3s)   
- 497664 s^2 e^{3s}\Ei_1(3s) +155520s^2 e^\frac{3s}{2}\Ei_1(\frac{3s}{2}) \\
\nonumber &+94976 + 675 s^5 e^\frac{3s}{2} +11097s^4 e^\frac{3s}{2} +15552  e^{3s} s^4 \Ei_1(\frac{3s}{2}) 
-31104 s^4 e^{3s}\Ei_1(3s)  
\bigg).
\end{align}
\end{widetext}
\end{document}